\documentclass[]{aa}  

\usepackage{graphicx}
\usepackage{natbib} 
\bibpunct{(}{)}{;}{a}{}{,} 

\usepackage{txfonts}
\usepackage{amssymb}
\def\gapprox{\;\rlap{\lower 2.5pt            
 \hbox{$\sim$}}\raise 1.5pt\hbox{$>$}\;}
\def\lapprox{\;\rlap{\lower 2.5pt            
 \hbox{$\sim$}}\raise 1.5pt\hbox{$<$}\;}

\begin{document}

\title{Quadrant polarization parameters for the scattered light of
      circumstellar disks.}

\subtitle{Analysis of debris disk models and observations of HR 4796A}

    \author{H.M.~Schmid}

\institute{
  ETH Zurich, Institute for Particle Physics and Astrophysics,
  Wolfgang-Pauli-Strasse 27, CH-8093 Zurich, Switzerland, \\
  \email{schmid@astro.phys.ethz.ch}}

           \date{Received ...; accepted ...}

 
\abstract
    {Modern imaging polarimetry provides spatially
      resolved observations for many circumstellar disks
      and quantitative results for the measured polarization
      which can be used for comparisons with
      model calculations and for systematic studies of disk samples.}
    {This paper introduces the quadrant
      polarization parameters $Q_{000}$, $Q_{090}$, $Q_{180}$,
      $Q_{270}$ for Stokes $Q$ and $U_{045}$, $U_{135}$, $U_{225}$, $U_{315}$
      for Stokes $U$  for circumstellar disks and describes their use for the
      polarimetric characterization of the dust in debris disks.}
    {We define the quadrant polarization parameters $Q_{xxx}$ and $U_{xxx}$
       and illustrate their properties with measurements of
       the debris disk around HR~4796A from \citet{Milli19}.
      We calculate quadrant parameters for simple models of rotationally
      symmetric and optically thin debris disks and the results
      provide diagnostic diagrams for the determination
      of the scattering asymmetry of the dust. This method is tested
      with data for HR~4796A and compared
      with detailed scattering phase curve extractions
      in the literature.}
      {The parameters $Q_{xxx}$ and $U_{xxx}$ are ideal
        for a well-defined and simple characterization of the azimuthal
        dependence of
        the polarized light from circumstellar disk because they
        are based on the ``natural'' Stokes $Q$ and $U$ quadrant pattern
        produced by circumstellar scattering.
        For optically thin and rotationally symmetric debris disks the quadrant
        parameters normalized to the integrated azimuthal polarization
        $Q_{xxx}/Q_\phi$ and $U_{xxx}/Q_\phi$ or quadrant ratios like
        $Q_{000}/Q_{180}$ depend only on
        the disk inclination $i$ and the polarized scattering phase function
        $f_\phi(\theta)$ of the dust, and they do not depend on the radial
        distribution of the scattering emissivity. Because the
        disk inclination $i$ is usually well known for resolved observations,
        we can derive
        the shape of $f_\phi(\theta)$ for the phase angle range $\theta$
        sampled by the polarization quadrants. This
        finding  also applies to models with vertical extensions as observed
        for debris disks. Diagnostic diagrams are calculated for
        all normalized quadrant parameters and
        several quadrant ratios for the determination
        of the asymmetry parameter $g$ of the polarized Henyey-Greenstein
        scattering phase function $f_\phi(\theta,g)$.
        We apply these diagrams to the measurement of HR~4796A, and find
        that a phase function with only one parameter does not reproduce
        the data well. We find a better solution with a three-parameter
        phase function $f_\phi(\theta,g_1,g_2,w)$,
        but it is also noted that the well-observed
        complex disk of HR~4796A cannot be described in full
        detail with the simple quadrant polarization parameters.}
      {The described quadrant polarization parameters are very
        useful for quantifying the azimuthal dependence of the scattering
        polarization of spatially resolved circumstellar disks illuminated
        by the central star. They provide a simple test of the
        deviations of the disk geometry from axisymmetry and 
        can be used to constrain the scattering phase
        function  for optically thin disks without detailed model
        fitting of disk images.
        The parameters are easy to derive from observations and
        model calculations and are therefore well suited to
        systematic studies of the dust scattering in circumstellar disks.}


\authorrunning{H.M. Schmid}

\titlerunning{Quadrant polarization parameters for circumstellar disks}

   \maketitle
%


\section{Introduction} \label{Introduction}

Circumstellar disks reflect the light from the central star,
and the produced scattered intensity and polarization
contain a lot of information about the disk geometry and
the scattering dust particles.
The scattered light of disks is usually only 
a contribution of a few percent or less to the direct light from
the central star and therefore requires observations
with sufficiently high resolution and contrast to resolve the disk
from the star. This was achieved in recent years for many
circumstellar disks with adaptive optics (AO) systems at large
telescopes using polarimetry, a powerful high-contrast
technique, to disentangle the scattered and
therefore polarized light of the disk from the direct and
typically unpolarized light of the central star
\citep[e.g.,][]{Apai04,Oppenheimer08,Quanz11,Hashimoto11,Muto12}.

With AO systems, the observational point spread function (PSF)
depends to a large extent on the atmospheric turbulence and is highly
variable \citep[e.g.,][]{Cantalloube19}. For this reason, the disks are often
only detected in polarized light and it is not possible to disentangle the
disk intensity signal from the strong, variable PSF of the central
star \citep[see e.g.,][]{Esposito20}.
Therefore, analyses of the scattered light from the disk are often
based on the differential polarization alone and only in favorable cases
can one combine this with measurements of the disk intensity.
For the data analysis, the observed polarization must first be corrected for
instrumental polarization effects, and this is relatively difficult for
complex AO systems \citep[e.g.,][and references therein]{Schmid21}.
For this reason, the first generation of AO systems with polarimetric mode
provided useful qualitative polarimetric
results but hardly any quantitative results.

This situation has changed with the new extreme AO
systems GPI \citep{Macintosh14} and SPHERE \citep{Beuzit19},
which, in addition to better image quality, also  provide polarimetrically calibrated
data for the circumstellar disk \citep{Perrin15,Schmid18,deBoer20,vanHolstein20}.
Thus, quantitative polarization measurements are now possible
for many circumstellar disks. However, the technique is not yet well established,
and detailed studies have only been made for a few bright,
extended disks; for example for HR 4796A \citep{Perrin15,Milli19,Arriaga20},
HIP 79977 \citep{Engler17}, HD 34700A \citep{Monnier19},
HD 169142 \citep{Tschudi21}, and
HD 142527 \citep{Hunziker21}.
There  also exist a few polarimetric studies based
on HST polarimetry, such as those for the disks of AU Mic \citep{Graham07}
or AB Aur \citep{Perrin09}. The large majority of polarimetric
disk observations in the literature focus their analysis on the
high-resolution disk geometry \citep[e.g.,][]{Benisty15,Garufi16,Avenhaus18},
and therefore the measurements are not calibrated and polarimetric
cancelation effects introduced by the limited spatial resolution
are not taken into account.
Even for the detailed polarimetric studies mentioned above,
the derived photo-polarimetric parameters are relatively heterogeneous.
Convolution effects are only sometimes taken into account,
and measurements are given as observed maps, azimuthal or radial profiles,
or as dust parameters of a well-fitting disk model.
This makes a comparison of results between
different studies difficult and inaccurate, in particular
because measuring uncertainties and model ambiguities are
rarely discussed in detail. 

Therefore, this paper introduces Stokes $Q$ and $U$
quadrant polarization parameters, which are simple to
derive but still well defined,
and facilitate systematic, quantitative studies of larger samples
of circumstellar disks
in order to make polarimetric measurements more valuable for
our understanding of the physical properties of the scattering dust.
Quantitative polarimetry of disks might be very useful to clarify
whether dust properties are different or similar for systems with
different morphology, age, level of illumination,
or dust composition.
The quadrant polarization parameters should also be useful for model
simulations describing the appearance of a given disk
for different inclinations so that intrinsic properties can be
disentangled from the effects of a particular disk inclination.
The same parameters can also be used to quantify the impact
of the convolution of the intrinsic disk signal or of simulated
images with instrumental PSF profiles to correct the observable
polarization of small and large disks for the effects of smearing and polarimetric degradation \citep{Schmid06}.

Using Stokes $Q$ and $U$ quadrant parameters for the
description of the polarization of disks is a new approach and
to the best of our
knowledge this is the first publication  using this type of analysis.
These quadrant polarization parameters 
are introduced in Sect.~2 using the published data of
\citet{Milli19} for the bright debris disk HR 4796A  as an example.
The interpretation of the measured values is illustrated with corresponding
model calculations for optically
thin debris disks, which are described in Sect.~3. The models
are similar to the classical simulation for the
scattered intensity of debris disks of \citet{Artymowicz89}
and \citet{Kalas96}, but they also consider the scattering
polarization as in the models of \citet{Graham07} and
\citet{Engler17}.
It is shown in Sect.~4 that the relative quadrant polarization
values for optically thin, axisymmetric (and flat) debris disks
are independent of the radial dust distribution and they depend
only on the disk inclination $i$ and the shape of the polarized
scattering phase function $f_\phi(\theta)$. Therefore,
in Sect.~5 we construct simple diagnostic diagrams
for quadrant polarization parameters which constrain
the $f_\phi(\theta)$ function for a given $i$, or 
directly yield the scattering asymmetry parameter $g$ if we adopt
a Henyey-Greenstein scattering function for the dust.  
The diagnostic diagrams are applied to the quadrant polarization
measurements of HR 4796A and the obtained results are compared with
the detailed, model-free
phase-curve extraction of \citet{Milli19}.
Finally, in Sect.~6 we discuss the potential
and limitations of the quadrant polarization parameters
for the analysis of disk observations and model simulations in a broader context.

\section{Quadrant polarization parameters for disks}

\subsection{Polarization parameters in sky coordinates}
Polarimetric imaging of stellar systems with
circumstellar disks provides typically sky images for the intensity  
$I_{\rm obs}(\alpha,\delta)$ and the Stokes linear polarization parameters
$Q_{\rm obs}(\alpha,\delta)$ and $U_{\rm obs}(\alpha,\delta)$. For dust scattering,
the circular polarization is expected to be much smaller than the
linear polarization and is usually not measured; it is therefore neglected in this work. $Q$ and $U$ are differential quantities
for the linear polarization components
\begin{equation}
  Q=I_0-I_{90}\quad {\rm and}\quad U=I_{45}-I_{135}\,,
\end{equation}
which can also be expressed as polarization flux
$P=p\times I=(Q^2+U^2)^{1/2}$ and polarization position angle
$\theta_P= 0.5\,{\rm atan2}(U,Q)$\footnote{defined as in
  the FORTRAN function atan2$(y,x)$ for Cartesian to
polar coordinate conversions}. The polarized flux
$p\times I$ is by definition a positive quantity, which,  for
noisy $Q$ or $U$ imaging data, suffers from a significant bias effect
\citep{Simmons85}.

The azimuthal Stokes parameter $Q_\phi$ can be used as an alternative
for $P$ for circumstellar disks. $Q_\phi$ measures polarization in
the azimuthal direction with respect to the central star
$(\alpha_0,\delta_0)$ and $U_\phi$ in a direction
rotated by 45$^\circ$ with respect to azimuthal. Circumstellar dust,
which scatters light from the central star, 
mostly produces polarization with azimuthal orientations
$\theta_P\approx \phi_{\alpha\delta}+90^\circ$,
while $U_\phi$ is almost zero.
Therefore, $Q_\phi$ can
be considered as a good approximation for the polarized flux
of the scattered radiation from circumstellar disks
$p\times I =  (Q_\phi^2+U_\phi^2)^{1/2} \approx Q_\phi$,
and this approximation also avoids the noise bias
problem \citep{Schmid06}. For the model calculations of optically
thin disks presented in this work, there is strictly $p\times I=Q_\phi$,
while small signals $U_\phi< 0.05~Q_\phi$ can be produced by
multiple scattering in optically thick disks \citep{Canovas15} or disks
with aligned aspherical scattering particles. The
$U_\phi$ signal is often much larger, that is, $U_\phi\gapprox 0.1~Q_\phi$, in observations
because of the PSF convolution problem for poorly resolved disks
and polarimetric calibration errors. Both effects should be taken
into account and corrected for the polarimetric measurements of disks.
The azimuthal Stokes parameters are defined by
\begin{eqnarray}
Q_\phi(\alpha,\delta)=-Q(\alpha,\delta)\cos(2\phi_{\alpha\delta})-U(\alpha,\delta)\sin(2\phi_{\alpha\delta})\,,\\
U_\phi(\alpha,\delta)=+Q(\alpha,\delta)\sin(2\phi_{\alpha\delta})-U(\alpha,\delta)\cos(2\phi_{\alpha\delta})\,,
\end{eqnarray}
with
\begin{equation}
\phi_{\alpha\delta}={\rm atan2}((\delta-\delta_0),(\alpha-\alpha_0)) \,,
\end{equation}
according to the description of \citet{Schmid06} for the radial
Stokes parameters $Q_r$, $U_r$, and using $Q_\phi=-Q_r$ and $U_\phi=-U_r$. 

\subsection{Disk integrated polarization parameters}

Quantitative measurements of the scattered radiation
from circumstellar disks were obtained in the past
with aperture polarimetry, which provided the disk-integrated
Stokes parameters
$\overline{Q}$, $\overline{U}$ or the polarized flux
$\overline{P}=(\overline{Q}^2+\overline{U}^2)^{1/2}$
usually expressed as fractional polarization relative to
the system-integrated intensity $\overline{Q}/\overline{I}_{\rm tot}$,
$\overline{U}/\overline{I}_{\rm tot}$, or $\overline{P}/\overline{I}_{\rm tot}$
and the averaged polarization position angle
$\langle \theta_p \rangle=0.5\,{\rm atan2}(\overline{U},\overline{Q})$
\citep[e.g.,][]{Bastien82,Yudin98}.
This polarization signal can be attributed to the scattered
light from the disk if the star produces no polarization and
if the interstellar polarization can be neglected, or if these
contributions can be corrected. Usually, the disk intensity 
$\overline{I}=\overline{I}_{\rm tot}-\overline{I}_{\rm star}$
cannot be distinguished from the stellar intensity
with aperture polarimetry,
apart from a few exceptional cases, such as that of $\beta$ Pic.

Aperture polarimetry provides only the net scattering polarization,
but misses potentially strong positive and negative polarization
components $+Q,-Q$ and $+U,-U$ which cancel each other in unresolved
observations. Therefore, $\overline{Q}$ and $\overline{U}$, or $\overline{P}$
and $\langle \theta_p \rangle$ provide
only one value and one direction for circumstellar disks, which agglomerate all possible
types of deviations from axisymmetry of the scattering polarization of a
circumstellar disk.

Disk-resolved polarimetric imaging avoids or strongly
reduces the destructive cancellation effect and provides therefore
much more information about the scattering polarization of disks.
The most basic polarization parameter for the characterization
of a resolved disk is the disk integrated azimuthal polarization
$\overline{Q}_\phi$ which can be considered as equivalent to the
polarized flux for resolved observations $\overline{\cal{P}}$
\footnote{We distinguish between the polarized flux derived from
resolved observations $\overline{\cal{P}}$ and unresolved aperture
polarimetry $\overline{P.}$}. The integrated polarized flux
$\overline{Q}_\phi$ depends on the spatial resolution of the data; this aspect is not considered in this work because for well-resolved disks,
like that of HR 4796A, the effect of limited spatial resolution is small
and can be corrected with modeling of the instrumental smearing.

If the disk intensity $\overline{I}$ is measurable, one can also
determine the disk-averaged fractional polarization
$\langle p_\phi \rangle = \overline{Q}_\phi/\overline{I}$. 
Unfortunately, it is still often very difficult
to measure the disk intensity $I(\alpha,\delta)$ with AO-observations
because the signal cannot be separated from the intensity of the variable point
spread function $I_{\rm star}(\alpha,\delta)$ of the much brighter
central star. In these cases, the fractional polarization can only
be expressed relative to the total system intensity 
$\langle p_{\phi,{\rm tot}} \rangle = \overline{Q}_\phi/\overline{I}_{\rm tot}$.
Integrated or averaged quantities are well defined but 
not well suited to characterizing the polarimetric features and
the azimuthal dependence of the polarization for
circumstellar disks. Therefore, we introduce new polarization
parameters to quantify the individual
positive and negative polarimetric components $+Q,\,-Q$ and
$+U,\,-U$ of spatially resolved disk observations and models.

\begin{figure}
\includegraphics[width=8.8cm]{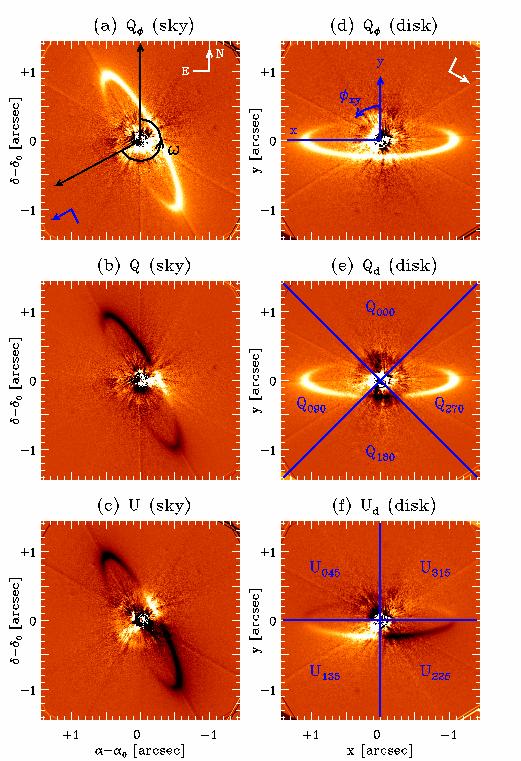}
\caption{Illustration of the definition of the quadrant polarization
  parameters (panels e and f) for the observations of the
    debris disk around HR 4796A from \citet{Milli19}. Left: 
    Azimuthal polarization $Q_\phi$ and Stokes $Q$ and $U$ in
    relative $\alpha,\delta$-sky coordinates; Right: $Q_\phi$, $Q_{\rm d}$,
    and $U_{\rm d}$ in $x,y$-disk coordinates.}
\label{FigHR4796A}
\end{figure}

\subsection{Quadrant polarization parameters in disk coordinates}

For polarimetric imaging of circumstellar disks, the geometric
orientation of the disk and the Stokes $Q$ and $U$ parameters are often
described in sky coordinates. This is inconvenient for the characterization of
the intrinsic scattering geometry of the disk and therefore
we define new polarization parameters $Q_{\rm d}$ and $U_{\rm d}$
in the disk coordinate
system $(x,y)$, where the central star is at $x_0=0,y_0=0$ and the
$x$ and $y$ are aligned with the major and minor axis of the projected
disk, respectively.
The positive $x$-axis is pointing left to ease the comparison
with observations in relative sky coordinates
$\alpha-\alpha_0$ and $\delta-\delta_0$ 
and to get the same convention for the $Q_{\rm d}$
and $U_{\rm d}$ orientations in $x,y$ and sky images.
Thus, the relative $\alpha-\alpha_0$, $\delta-\delta_0$ sky
coordinates of the observed
images $I$, $Q_\phi$, $Q,$ and $U$
must be rotated according to 
\begin{eqnarray}
  x = (\alpha-\alpha_0) \cos\omega + (\delta-\delta_0) \sin\omega\,, \\
  y = (\delta-\delta_0) \cos\omega - (\alpha-\alpha_0) \sin\omega \,.
\end{eqnarray}
as shown in Fig.~\ref{FigHR4796A} for the imaging polarimetry of
HR 4796A using $\omega=242^\circ$ ($-118^\circ$).
We use the convention that $\omega$ aligns the more distant semi-minor
axis of the projected disk with the positive (upward) $y$-axis.

Also, the Stokes parameters for the linear polarization must be rotated  
from the $Q,U$ sky system to the $Q_{\rm d},U_{\rm d}$ disk system
using the geometrically rotated $(x,y)$-frames
\begin{eqnarray}
  Q_{\rm d}(x,y) = Q(x,y)\,\cos (2 \omega)
       + U(x,y)\,\sin (2\omega) \,, \\  
  U_{\rm d}(x,y) = U(x,y)\,\cos (2 \omega)
       - Q(x,y)\,\sin (2\omega) \,. 
\end{eqnarray} 

We define for the $Q_{\rm d}(x,y)$ and $U_{\rm d}(x,y)$ polarization images
the quadrant parameters
$Q_{000}$, $Q_{090}$, $Q_{180}$, and $Q_{270}$ for Stokes $Q_{\rm d}$ and 
$U_{045}$, $U_{135}$, $U_{225}$, and $U_{315}$ for Stokes $U_{\rm d}$,
which are obtained  by integrating the Stokes $Q_{\rm d}$ or $U_{\rm d}$
disk polarization signal in the corresponding
quadrants as shown in Fig.~\ref{FigHR4796A} (e) and (f).
This selection of polarization parameters
is of course motivated by the natural $Q$ and $U$ quadrant
patterns for circumstellar scattering where the signal in a given
quadrant typically has the same sign   everywhere
and is almost zero at the borders of the
defined integration region. This is strictly the case for
all the model calculations for optically
thin debris disks presented in this work and the same type of
quadrant pattern is also predominant for the scattering polarization
of proto-planetary disks.
Multiple scattering and grain alignment effects can introduce
deviations from a ``clean'' quadrant polarization pattern which may
be measurable in high-quality observations \citep{Canovas15}. 
However, the smearing and polarization cancellation
effects introduced by the limited spatial resolution are typically  much
more important in affecting the quadrant pattern. For poor spatial
resolution or very small disks, the
quadrant pattern disappears \citep{Schmid06} or is strongly
disturbed for asymmetric systems \citep{Heikamp19}. 

Eight quadrant polarization values seems to be a useful number for
the characterization of the azimuthal distribution of the
polarization signal of disks. The parameters provide some
redundancy to check and verify systematic effects, or to allow alternative disk characterizations if one parameter
is not easily measurable or is affected by a special disk feature.
Let us consider the redundancy in the context of the geometrical symmetry
of disks and the dust scattering asymmetry.

For an inclined, but intrinsically axisymmetric disk,
the Stokes $Q_{\rm d}$ quadrants have the symmetry
\begin{equation}
    Q_{090} = Q_{270,}
\end{equation}
and the Stokes $U_{\rm d}$ quadrants have the anti-symmetries
\begin{equation}
  U_{045} = - U_{315}\quad {\rm and} \quad U_{135} = - U_{225}\,.
\end{equation}  
Special cases generate additional equalities; for example
the models with isotropic scattering (see Fig.~\ref{FigDisk2Di45}) have a
front--back symmetry and therefore there is also $Q_{000}=Q_{180}$ and
$U_{045}=-U_{135}=U_{225}=-U_{315}$. For an axisymmetric disk seen
pole-on, all quadrant parameters have the same absolute
value. If a disk deviates from an intrinsically symmetric
geometry, for example a brighter $+x$ side,
then this would result
in $|Q_{090}| > |Q_{270}|$ for Stokes $Q_{\rm d}$ and
$|U_{045}| > |U_{315}|$ or $|U_{135}| > |U_{225}|$ for Stokes $U_{\rm d}$.

Dust, which is predominantly forward scattering, produces
more signal for front side polarization quadrants compared to the backside quadrants and this is equivalent to $|Q_{180}|>|Q_{000}|$
or $|U_{135}|>|U_{045}|$ and $|U_{225}|>|U_{315}|$. Properties that can be deduced from parameter ratios derived from
the same data set are important because this reduces
the impact of at least some systematic uncertainties in the measurements. 

The quadrant parameters are also linked to the integrated
polarization parameters $\overline{Q}_\phi$, $\overline{Q}_{\rm d}$,
and $\overline{U}_{\rm d}$. For the sum of all four Stokes $Q_{\rm d}$ and Stokes $U_{\rm d}$
quadrants, there is
\begin{equation}
  {\Sigma\, Q_{xxx}} = \overline{Q}_{\rm d} \quad {\rm and}
              \quad {\Sigma\, U_{xxx}}=\overline{U}_{\rm d} \,.
\end{equation}
For pole-on systems, there is $\overline{Q}_{\rm d}=\overline{U}_{\rm d}=0$,
because of the symmetric cancellation of positive and negative
quadrants and intrinsically axis-symmetric systems have $\overline{U}_{\rm d}=0$
for all disk inclinations because of the left--right antisymmetry. 
Axisymmetric but inclined systems have 
$\overline{Q}_{\rm d}\neq 0$ in general. For disks with larger
inclination $i,$ a smaller fraction
of the disk is ``located'' in the quadrants $Q_{000}$ and $Q_{180}$
and a larger fraction is located in $Q_{090}$ and $Q_{270}$ near the major axis
because of the disk projection. Edge-on disks are almost only
located in the quadrants $Q_{090}$ and $Q_{270}$ and quadrant
sums approach $Q_{000}+Q_{180}\rightarrow 0$ and
$Q_{090}+Q_{270}\rightarrow \overline{Q}_{\rm d}$ for $i\rightarrow 90^\circ$.

For the sums of the four absolute quadrant parameters for Stokes $Q_{\rm d}$,
there is of course
\begin{equation}
\Sigma\,|Q_{xxx}| > \Sigma\, Q_{xxx}\quad {\rm and}\quad \Sigma\,|Q_{xxx}|<\overline{Q}_\phi,
\end{equation}
and the equivalent exists for sums of the Stokes $U_{\rm d}$ quadrants.
For a pole-on view, the system
is axisymmetric with respect to the line of sight and there is
\begin{equation}
\Sigma\, |Q_{xxx}| = \Sigma\, |U_{xxx}| =\frac{2}{\pi}\,\overline{Q}_\phi\,,
\end{equation}
where each quadrant has the same absolute value
of $\overline{Q}_\phi/2\pi=0.159\,\overline{Q}_\phi$.
These sums will converge for edge-on disks $i=90^\circ$
to $\Sigma\, |Q_{xxx}|\rightarrow \overline{Q}_\phi$ for Stokes $Q_{\rm d}$ and
to $\Sigma\, |U_{xxx}|\rightarrow 0$ for Stokes $U_{\rm d}$.

\begin{figure}
\includegraphics[width=8.8cm]{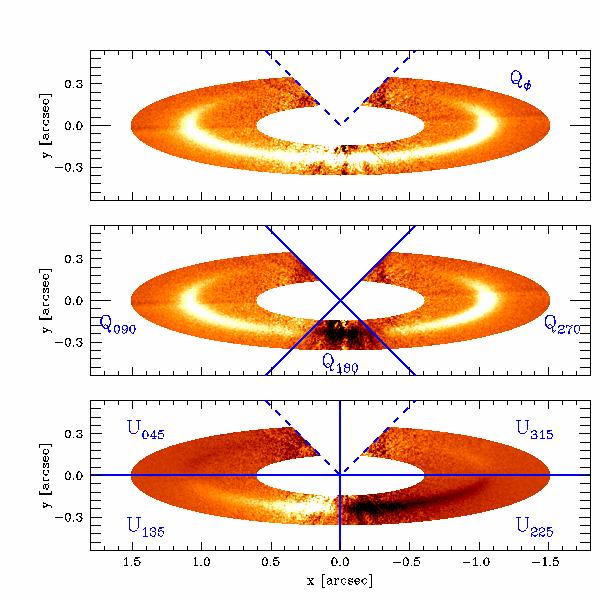}
\caption{Apertures used for the measurements of the
  integrated azimuthal polarization $\overline{Q}_\phi$ (top),
  the Stokes $Q_{\rm d}$ quadrants (middle), and the Stokes $U_{\rm d}$ quadrants
  (bottom) for HR~4796A.}
\label{FigApertureHR}
\end{figure}

\subsection{Quadrant polarization parameters for HR 4796A}
\label{Sect4796A}
We use the high-quality differential polarimetric imaging (DPI) 
of the bright debris disk HR 4796A from \citet{Milli19}
shown in Fig.~\ref{FigHR4796A} as an example for the
measurement of the quadrant polarization parameters.
The data were taken with the SPHERE/ZIMPOL instrument
\citep{Beuzit19,Schmid18}
in the very broad band (VBB) filter with the central wavelength
$\lambda_c=735$~nm and full width of $\Delta\lambda = 290~$nm.

These data provide ``only'' the differential 
Stokes $Q(x,y)$ and $U(x,y)$ signals or the corresponding azimuthal
quantities $Q_\phi(x,y)$ and $U_\phi(x,y)$, but no intensity signal
because it is difficult with AO-observations to separate the disk
intensity from the variable intensity PSF of the much brighter
central star. The scattered light of the disk around
HR 4796A was previously detected in polarization and intensity with AO systems
in the near-IR \citep[e.g.,][]{Milli17,Chen20,Arriaga20}, and in intensity
in the visual with HST \citep[e.g.,][]{Schneider09}. 

Quadrant polarization parameters $Q_{xxx}$ and
$U_{xxx}$
for HR 4796A derived from the data shown in Fig.~\ref{FigHR4796A}(e) and (f)
are given in Table~\ref{QuadHR4796A} as relative values using the
integrated polarized flux $\overline{Q}_\phi$ as reference.
The quadrant values were obtained by integrating the counts in the
annular apertures sections as illustrated in Fig.~\ref{FigApertureHR},
which avoid the high noise regions from the PSF peak in the center.
The uncertainties indicated in Table~\ref{QuadHR4796A}
account for the image noise, but do not account
for systematic effects related to the selected aperture geometry or
polarimetric calibration uncertainties. The noise errors are particularly
large for the quadrants $Q_{000}$ and $Q_{180}$ because of the
small separation of these disk sections from the bright star and
additional negative noise spikes which are particularly strong for the
$Q_{000}$ quadrant. Because of this noise, the formal integration
gives $Q_{000}/\overline{Q}_\phi=-0.04$. From the decreasing trend
of the signal in for example $Q_\phi$ towards the back side of the disk and
the measured disk signals of about 0.03 for
$U_{045}/\overline{Q}_\phi$ or $U_{315}/\overline{Q}_\phi$, an 
absolute signal of less than $|Q_{000}|/\overline{Q}_\phi<0.01$
is expected for a smooth dust distribution in the ring and
any reasonable assumptions for the dust scattering. Therefore,
we do not consider the noisy $Q_{000}$-measurement in the quadrant
sums $\Sigma Q_{xxx}$. The noise in the $Q_{180}$-quadrant is also
significantly enhanced when compared to other
quadrants despite the relatively strong signal and the small integration
area. A detailed analysis of the noise pattern might improve
the measuring accuracy, but this is beyond the scope of this paper.

Because of the noise, the apertures for $Q_\phi$ and the quadrants
$U_{045}$ and $U_{135}$ were restricted
to avoid the noisy and essentially signal-free region around
$\phi_{xy}\approx 0^\circ$. One should note that uncertainties
for the quadrant measurements, for example
for $Q_{180}$,  also affect the $Q_{\phi}$ value and a dominant
noise feature therefore  has an enhanced impact on relative parameters
such as $Q_{xxx}/\overline{Q}_\phi$, which include the disk
integrated azimuthal polarization $\overline{Q}_\phi$. These are
typical problems for high-contrast observations of inclined
disks and it can be very useful to select only high-signal-to-noise
quadrant values for the characterization of the azimuthal
dependence of the disk polarization.

\begin{table}
  \caption{Measured relative quadrant polarization
    parameters for HR 4796A, deviations $\Delta$ (in $\%$) from left--right
    symmetry and ratios $\Lambda$ for back--front flux ratios.}
\label{QuadHR4796A}
\begin{tabular}{lclc}
\hline \hline
\noalign{\smallskip}
\multispan{2}{\hfil Stokes $Q_{\rm d}$ quadrants\hfil}
& \multispan{2}{\hfil Stokes $U_{\rm d}$ quadrants\hfil}\\
\noalign{\smallskip\hrule\smallskip}
\noalign{relative quadrant polarization (errors: $\approx \pm 0.010$)}  
$Q_{000}/\overline{Q}_\phi$ & $-0.04$\tablefootmark{a} & $U_{045}/\overline{Q}_\phi$ & $-0.031$  \\
$Q_{090}/\overline{Q}_\phi$ & $+0.385$    & $U_{135}/\overline{Q}_\phi$ & $+0.212$  \\
$Q_{180}/\overline{Q}_\phi$ & $-0.076$\tablefootmark{b}
                             & $U_{225}/\overline{Q}_\phi$ & $-0.159$  \\
$Q_{270}/\overline{Q}_\phi$ & $+0.343$    & $U_{315}/\overline{Q}_\phi$ & $+0.033$  \\
\noalign{\smallskip}
\noalign{quadrant sums (errors: $\approx \pm 0.030$) }
$\Sigma Q_{xxx}/\overline{Q}_\phi$ & $+0.652$\tablefootmark{c}  & $\Sigma U_{xxx}/\overline{Q}_\phi$ & $+0.055$ \\ 
$\Sigma|Q_{xxx}|/\overline{Q}_\phi$ & $0.804$\tablefootmark{c}  & $\Sigma|U_{xxx}|/\overline{Q}_\phi$ & $0.435$ \\ 
\noalign{\smallskip}
\noalign{left--right asymmetry parameters}
\noalign{\smallskip}
$\Delta^{090}_{270}$ &~~~~~$+6\pm 2~\%$~~~~~& $\Delta^{135}_{225}$ & $+14\pm 4~\%$ \\
\noalign{\smallskip}
                   &                & $\Delta^{045}_{315}$ & $-3\pm 23~\%$ \\
\noalign{\smallskip}
\noalign{back--front parameter ratios}
\noalign{\smallskip}
$\Lambda^{000}_{180}$ & $<0.6$   & $\Lambda^{045}_{135}$ & $0.15\pm 0.07$ \\
\noalign{\smallskip}
                    &          & $\Lambda^{315}_{225}$ & $0.21\pm 0.09$ \\
\noalign{\smallskip}
\noalign{special back--front parameter ratios}
\noalign{\smallskip}
\multispan{3}{$\Lambda_a=(|Q_{090}|+|Q_{270}|)/(2\,|Q_{180}|)$\hfil}
                   & $4.8 \pm 0.7$ \\
\noalign{\smallskip}
\multispan{3}{$\Lambda_b=(|Q_{090}|+|Q_{270}|)/(|U_{135}|+|U_{225}|)$\hfil}
                   & $2.0 \pm 0.2$ \\
\noalign{\smallskip\hrule\hrule\smallskip}
\end{tabular}
\tablefoot{
\tablefoottext{a}{strongly affected by noise, the expected signal is
  $|Q_{000}|<0.01$;}
\tablefoottext{b}{errors: $\approx \pm 0.020$;}
\tablefoottext{c}{$Q_{000}$ is not included in $\Sigma Q_{xxx}$.}}
\end{table}

Asymmetries between the left and right or the positive and negative sides
of the $x$-axis can be deduced from the absolute quadrant parameters
$|Q_{090}|$ and $|Q_{270}|$, $|U_{045}|$ and
$|U_{315}|$, and $|U_{135}|$ and $|U_{225}|$.
Table~\ref{QuadHR4796A} gives relative left--right brightness differences
$\Delta^{aaa}_{bbb}$ calculated according to
\begin{equation}
  \Delta^{090}_{270} = \frac{|Q_{090}|-|Q_{270}|}{|Q_{090}|+|Q_{270}|}
\end{equation}
for Stokes $Q_{\rm d}$ parameters and equivalent for $\Delta^{045}_{315}$
and $\Delta^{135}_{225}$ for Stokes $U_{\rm d}$. 

The asymmetry values $\Delta^{090}_{270}$ and $\Delta^{135}_{225}$ 
both yield more flux on the left side of the $(x,y)$-plane, which is the SW-side for the HR 4796A disk.
This does not agree with previous determinations, including 
even the analysis of the same data by \citet{Milli19} who
measured more flux on the NE side.
A more detailed investigation reveals that the peak surface
brightness is indeed higher for the disk on the NE side and that
the left--right asymmetry depends on
the width of the annular apertures used for the flux extraction.
If we integrate only a narrow annular region with a full width of
$\Delta x = 0.1''$ at the location of the
major axis then we also get more flux for the NE side or negative
$\Delta^{090}_{270}$-values of $-3\pm 2~\%$ but still less than
the ``negative'' (SW-NE) asymmetry measured by \citet{Milli19} for
$Q_\phi$ or \citet{Schneider18} for the intensity. We therefore conclude
that there are subtle asymmetries at the level of $\Delta\approx 10~\%$ present
for the disk HR 4796A, which are positive for a wide flux extraction
and negative for a narrow flux extraction. Measurement uncertainties in the
left--right differences are at the few percent level for bright
quadrant pairs. 

The back-side to front-side brightness contrast can be expressed with
quadrant ratios such as\begin{equation}
  \Lambda^{000}_{180} = \frac{|Q_{000}|}{|Q_{180}|}\,, \ \ {\rm and} \ \
  \Lambda^{045}_{135} = \frac{|U_{045}|}{|Q_{135}|}\,,
  \ \ {\rm or} \ \ \Lambda^{315}_{225} = \frac{|U_{315}|}{|U_{225}|}\,,
\end{equation}
and their equivalents for other quadrant ratios. These ratios
are small $\lapprox 0.5$ for HR 4796A which is indicative of dust with
a strong forward scattering phase function. For isotropic scattering
in an axisymmetric disk, the ratios would be
$\Lambda^{000}_{180}=\Lambda^{045}_{135}=1$ for all inclinations. 
Because the polarization flux in the backside quadrants is small, one
can also assess the disk forward scattering with a
comparison of the brighter $Q_{090}$ and $Q_{270}$ quadrants with
the Stokes $Q_{\rm d}$ front quadrant $Q_{180}$ or the Stokes $U_{\rm d}$ front quadrants
$U_{135}$ and $U_{225}$ as given in Table~\ref{QuadHR4796A}.

As demonstrated in Table~\ref{QuadHR4796A}, the polarization quadrants
provide a useful set of parameters for the quantitative characterization
of the geometric distribution of the polarization signal in
circumstellar disks. Comparisons with model calculations are required
to assess the diagnostic power of the derived values for
the determination of the scattering properties of the dust or
for the interpretation of the strength of disk asymmetries.


\section{Disk model calculations}
\label{SectModel}

The simple disk models presented in this work
follow the basic calculations
for the scattered intensity from debris disks \citep{Artymowicz89,Kalas96}
and the scattering polarization
\citep[e.g.,][]{Bastien88,Whitney92,Graham07,Engler17} but focus on the
azimuthal dependence of the scattering polarization and the determination of
the quadrant polarization values. The model disks are described
by a dust-density distribution
in cylindrical coordinates $\rho(r,\varphi_d,h)$ where $r$ is the
radius vector $r=(x_d^2+y_d^2)^{-1/2}$ and $\varphi_d$ the azimuthal
angle in the disk plane\footnote{the azimuthal angle
$\varphi_d$ defined in the disk plane is different from the
azimuthal angle $\phi_{xy}$, which is defined in the sky plane.},
where $x_d$ coincides
with $x$ for the major axis of the projected, inclined disk. 
We consider in this work very simple, optically thin $\tau\ll 1$,
axisymmetric models $\rho(r,h)$, where
the dust scattering cross-section per unit mass $\sigma_{\rm sca}$
is independent of the location. The total dust scattering emissivity 
(integrated over all directions) $\epsilon(r,h)$
of a volume element is given by the stellar flux
$F_\lambda(r,h)=L_\lambda/(4\pi (r^2+h^2))$, the
disk density $\rho(r,h)$, and $\sigma_{\rm sca}$:
\begin{equation}
\epsilon(r,h) = \frac{L_\lambda}{4\pi (r^2+h^2)}\, \rho(r,h)\,\sigma_{\rm sca}\,.
\end{equation}
The incident flux decreases as $F_\lambda\propto 1/R^2$ with $R^2=r^2+h^2$,
because in our optically thin scattering model we neglect
the extinction of stellar light by the dust and the addition
of diffuse light produced by the scatterings in the disk.

The resulting images for the scattered light intensity $I(x,y)$
and the azimuthal polarization $Q_\phi(x,y)$ are obtained with a
line of sight or z-axis integration of the scattering emissivity
$\epsilon$ and the scattering phase functions $f_I(\theta_{xyz},g)$
for the intensity 
\begin{equation}
  I(x,y) = \int  \epsilon(x,y,z)\,f_I(\theta_{xyz},g)\, {\rm d}z \,,
\label{EqintI}  
\end{equation}  
and $f_\phi(\theta_{xyz},g,p_{\rm max})$ for the polarized intensity
\begin{equation}
  Q_\phi(x,y) = \int \epsilon(x,y,z) f_\phi(\theta_{xyz},g,p_{\rm max})\, {\rm d}z \,.
\label{EqintQ}  
\end{equation}
The transformations from the disk coordinate system ($r,\varphi_d,h$)
with $x_d=r \sin\varphi_d$ and $y_d=r \cos\varphi_d$ 
to the sky coordinate system $(x,y,z)$ is given by
$x = x_d$, $y = y_d \cos i+h\sin i$ and
$z = y_d \sin i - h\cos i$
where $i$ is the disk inclination. This also defines the scattering
angle $\theta_{xyz}$ and the radial separation to the central star $R_{xyz}$
for each point (x,y,z).

\subsection{Scattering phase functions}
The scattering phase function $f_I(\theta,g)$
for the intensity is described by the Henyey-Greenstein function
or HG-function \citep{Henyey41}, where $\theta$ is the angle of deflection 
\begin{equation}
  f_I(\theta,g) = \frac{1}{4\pi} \,
  \frac{1-g^2}{(1+g^2-2g\cos\theta)^{3/2}} \,.
\label{EqScatPhase}
\end{equation}
The asymmetry parameter $g$ is defined between $-1$ and $+1$
and backward scattering dominates for negative $g$,
forward scattering dominates for positive $g$, while the scattering is
isotropic for $g=0$.

The scattering phase function $f_\phi(\theta,g,p_{\rm max})$
for the polarized flux adopts the same angle dependence for the
fractional polarization as Rayleigh scattering, but with the
scale factor $p_{\rm max}\leq 1$
for the maximum polarization at $\theta=90^\circ$. This 
description is often used \citep[e.g.,][]{Graham07,Buenzli09,Engler17}
as a simple approximation for a Rayleigh scattering-like
angle dependence but reduced polarization induced by dust
particles \citep[e.g.,][]{Kolokolova10,Min16,Tazaki19}.

The angle dependence
of the fractional polarization of the scattered light is
\begin{equation}
  p_{\rm sca}(\theta,p_{\rm max}) = \frac{Q_{\rm sca}}{I_{\rm sca}}
  =p_{\rm max}\,\frac{1-\cos^2\theta}{1+\cos^2\theta} \,.
  \label{Eqpsca}
\end{equation}
The scattered intensity $I_{\rm sca}$ can be
split into the perpendicular $I_\perp$ and parallel $I_\parallel$
polarization components with respect to the scattering plane,
so that $I_{\rm sca}=I_\perp+I_\parallel$,
$Q_{\rm sca} = I_\perp-I_\parallel$ and $I_\perp = (I_{\rm sca} + Q_{\rm sca})/2$,
$I_\parallel = (I_{\rm sca} - Q_{\rm sca})/2$. Together with 
$p_{\rm sca}$, this yields 
\begin{eqnarray}
  I_\perp(p_{\rm max},\theta) =
  I_{\rm sca}\cdot \Bigl[0.5 +
  p_{\rm max}\Bigl(\frac{1}{1+\cos^2\theta}-0.5\Bigr)\Bigr]\,,
\label{Eqper} \\
  I_\parallel(p_{\rm max},\theta) = 
  I_{\rm sca}\cdot \Bigl[0.5 +
  p_{\rm max}\Bigl(\frac{\cos^2\theta}{1+\cos^2\theta}-0.5\Bigr)\Bigr]\,,
\label{Eqpar}
\end{eqnarray}
or, expressed as scattering phase functions,
$f_\perp=f_I\,k_\perp\,(\theta,p_{\rm max})$ and 
$f_\parallel=f_I\,k_\parallel\,(\theta,p_{\rm max})$, where
$k_\perp$ and $k_\parallel$ are the expressions
in the square brackets and $p_{\rm sca}=k_\perp-k_\parallel$.

\begin{figure}
\includegraphics[trim=2.0cm 12.5cm 3.0cm 3.0cm,width=8.8cm]{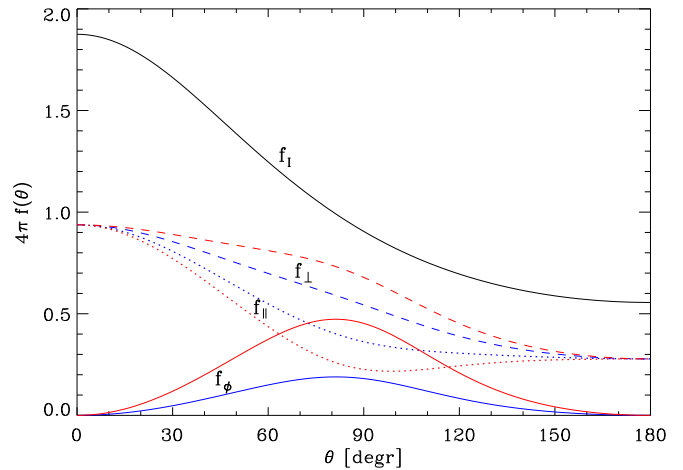}
\caption{Scattering phase functions for the HG asymmetry
  parameter $g=0.2$ for the total intensity $4\pi\,f_I(\theta)$ (black),
  for the polarized intensities $4\pi\,f_\perp(\theta)$ (dashed)
  and $4\pi\,f_\parallel(\theta)$
  (dotted), for the two values $p_{\rm max}=0.5$ (red)
  and $0.2$ (blue), and the corresponding scattering phase functions for the
  polarized flux $4\pi\,f_\phi(\theta)$
  (full, colored lines).}
\label{Figphasefunction}
\end{figure}

The scattering plane in a projected image of a circumstellar
disk has always a radial orientation with respect to the central star.
Therefore the induced scattering polarization $Q_{\rm sca}$, which is
perpendicular to the scattering plane, translates into an
azimuthal polarization $Q_\phi$ for the projected disk map.
The scattering phase functions are related by
\begin{equation}
  f_\phi(\theta,g,p_{\rm max}) = p_{\rm max}\,f^n_\phi(\theta,g)
      = f_I(\theta,g)\,p_{\rm sca}(\theta,p_{\rm max})\,,
\end{equation}
where $f_\phi$ can be separated into a normalized part $f^n_\phi$ for
$p_{\rm max}=1$ and the scale factor $p_{\rm max}$.

Figure~\ref{Figphasefunction} illustrates the scattering
phase functions for $g=0.2$ for the total intensity $f_I$
and the corresponding polarization components $f_\perp$ and $f_\parallel$
for $p_{\rm max}=0.5$ and 0.2. The differential phase function
$f_\phi=f_\perp-f_\parallel$ has the same
$\theta$-dependence $f_\phi^n(\theta,g)$ for both cases, and only the amplitude scales
with $p_{\rm max}$. In this formalism, the total intensity phase function
$f_I(g,\theta)$ does not depend
on the adopted polarization parameter $p_{\rm max}$. Expected values for
approximating dust scattering with HG scattering functions
are $p_{\rm max}\approx 0.05 - 0.8$, but we often set this scale
factor in this work  to $p_{\rm max}=1$ because this allows us to
plot the intensity and polarization on the same scale. A value
$p_{\rm max}=1$ applies for Rayleigh scattering but the HG-function
with $g=0$ (isotropic scattering) differs from the Rayleigh
scattering function for the intensity\footnote{Rayleigh scattering
produces more forward and backward scattering when compared to
isotropic scattering}. Hereafter, $f_\phi(\theta,g,p_{\rm max})$
and $f_\phi^n(\theta,g)$ are also referred to as the HG$_{\rm pol}$-function and
normalized HG$_{\rm pol}$-function, respectively.

The polarimetric scattering phase function for the
azimuthal Stokes parameters $f_\phi$ 
can be converted into phase functions $f_Q$ and $f_U$ for the Stokes
$Q_{\rm d}$ and $U_{\rm d}$ parameters 
\begin{equation}
  f_Q(\theta_{xyz},g,p_{\rm max}) =
       -f_\phi(\theta_{xyz},g,p_{\rm max})\,\cos(2\phi_{xy}), 
       \label{EqQQphi}
\end{equation}
\begin{equation} 
f_U(\theta_{xyz},g,p_{\rm max}) =
   -f_\phi(\theta_{xyz},g,p_{\rm max})\,\sin(2\phi_{xy}), 
\label{EqUQphi}
\end{equation}
where $\phi_{xy}={\rm atan2}(y,x)$  is the azimuthal angle in the
sky coordinates aligned with the disk as illustrated in Fig.~\ref{FigHR4796A}. 
These simple relations are valid in our optical thin (single scattering)
models because $U_\phi(x,y)=0$ and the Stokes $Q_{\rm d}(x,y)$ and $U_{\rm d}(x,y)$ model
images can then be calculated as in $Q_\phi$ in Eq.~\ref{EqintQ} but
using the phase functions for $f_Q$ and $f_U$.


\subsection{Flat disk models and azimuthal phase functions}

The calculations for the scattered flux $I(x,y)$ and
polarization $Q_\phi(x,y)$ with Eqs.~\ref{EqintI} and
\ref{EqintQ} are strongly simplified for a flat disk
because the integrations for a given
$x$-$y$ coordinate along the $z$-coordinate can be replaced
by single values for the separation $r_{xy}$ from the star,
for the scattering emissivity $\epsilon(r_{xy})$, and for the
scattering angle $\theta_{xy}$. The volume density
$\rho$ must be replaced by a vertical
surface density $\Sigma(r)$ or a line-of-sight surface density
$\Sigma(r)/\cos i$. Of course, the scattering in the disk
plane must still be treated as in an optically thin disk,
\begin{equation}
\tau(r) = \int_0^r\kappa_\Sigma\,\Sigma(r)\, {\rm d}r \ll 1\,,
\end{equation}
where $\kappa_\Sigma=a_\Sigma+\sigma_\Sigma$ is the disk extinction coefficient
composed of the contributions from absorption $a_\Sigma$ and scattering
$\sigma_\Sigma$.

\subsubsection{Projected flat disk image}
The scattered intensity and polarization for a flat disk are given by
\begin{eqnarray}
&&\hspace{-0.7cm} I(x,y) = \epsilon(r_{xy}) \,f_I(\theta_{xy},g), \\
&&\hspace{-0.7cm} Q_\phi(x,y) = \epsilon(r_{xy}) \,f_\phi(\theta_{xy},g,p_{\rm max}),
\end{eqnarray}  
and similar for the Stokes $Q_{\rm d}(x,y)$ and $U_{\rm d}(x,y)$ using
the phase function from Eqs.~\ref{EqQQphi} and \ref{EqUQphi}.
The scattering emissivity
\begin{equation}
\epsilon(r_{xy})= \frac{L_\lambda}{4\pi\,r_{xy}^2}\sigma_\Sigma\,\frac{\Sigma(r_{xy})}{\cos i}\,
\end{equation}
is proportional to the line-of-sight
surface density $\Sigma(r_{xy})/{\rm cos}\, i$, which considers the
disk inclination, and
where $r_{xy}^2 = x^2 + (y/\cos i)^2$ and
$\theta_{xy} ={\rm acos}(-z_{xy}/r_{xy}) = {\rm acos}(-y \tan i/r_{xy})$,
because $y=y_d \cos i$ and $z_{xy}=y_d \sin i$.

\begin{figure}
\includegraphics[width=9.5cm]{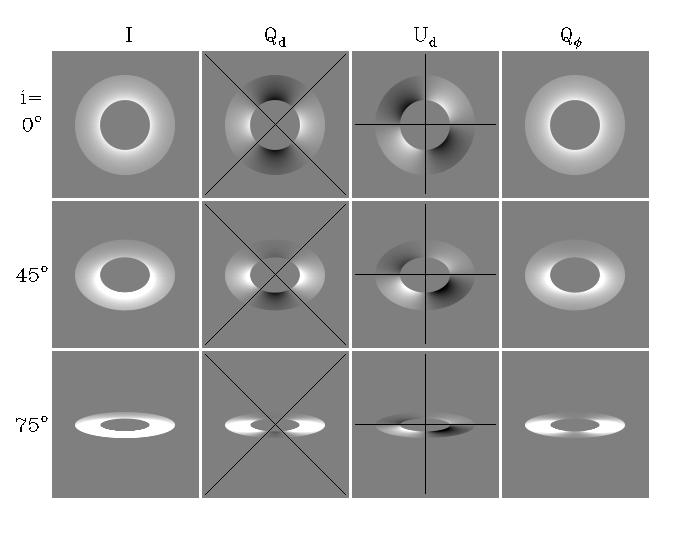}
\caption{$I$-, $Q_{\rm d}$-, $U_{\rm d}$- and $Q_\phi$-images for a flat disk with
  $g=0.3$, $p_{\rm max}=1$, and for inclinations $i=0^\circ$, $i=45^\circ$,
  and $75^\circ$. The same gray scale from $+a$ (white) to $-a$
  (black) is used for all panels. The black lines in the $Q_{\rm d}$ and $U_{\rm d}$
  panels indicate the polarization quadrants.}
\label{FigDisk2Dg3} 
\end{figure}

\begin{figure}
\includegraphics[width=9.5cm]{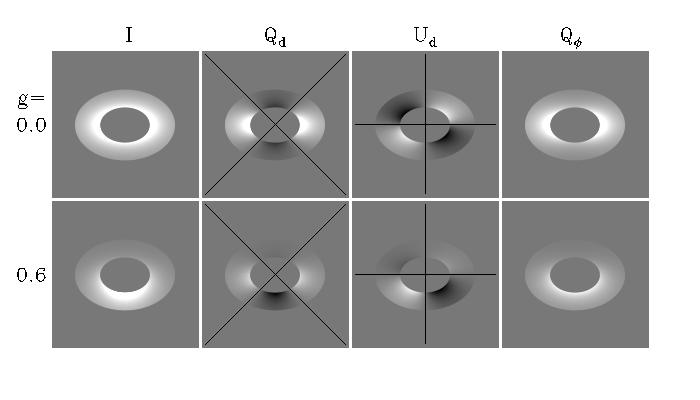}
\caption{$I$-, $Q_{\rm d}$-, $U_{\rm d}$- and $Q_\phi$-images for flat disks with
  $i=45^\circ$, $p_{\rm max}=1$, and for scattering asymmetry parameter
  $g=0.0$ and $0.6$. The same gray scale from $+a$ (white) to $-a$
  (black) is used for all panels.}
\label{FigDisk2Di45} 
\end{figure}

In these equations for $I(x,y)$, $Q_\phi(x,y)$, $Q_{\rm d}(x,y),$
and $U_{\rm d}(x,y),$ the radial
dependence of the scattering emissivity $\epsilon(r_{xy})$ is separated from
the azimuthal dependence described by the
scattering phase functions $f_I(\theta_{xy})$, $f_\phi(\theta_{xy})$, $f_Q(\theta_{xy}),$ and $f_U(\theta_{xy})$.
This is very favorable for the introduced quadrant parameters,
which describe the polarization of the scattered
light of disks with an azimuthal splitting of the signal 
$Q_{\rm d}(x,y)\rightarrow Q_{000}, Q_{090}, Q_{180}, Q_{270}$ and
$U_{\rm d}(x,y)\rightarrow U_{045}, U_{135}, U_{225}, U_{315}$ by integrating
the polarization in the corresponding quadrants outlined
by the black lines in the $Q_{\rm d}$ and $U_{\rm d}$ panels of
Figs.~\ref{FigDisk2Dg3} and \ref{FigDisk2Di45}.

Disk images for $I(x,y)$, $Q_{\rm d}(x,y)$, $U_{\rm d}(x,y)$, and $Q_\phi(x,y)$
for flat disk models are shown in Fig.~\ref{FigDisk2Dg3} for the asymmetry
parameter $g=0.3$, the scale factor $p_{\rm max}=1$,
and three different inclinations
$i=0^\circ,\, 45^\circ$, and $75^\circ$. For the radial surface
scattering emissivity, a radial dependence $\epsilon(r) \propto 1/r$
is adopted extending from an inner radius $r_1$ to the outer
radius $r_2=2\,r_1$. Along the
$x$-axis, the scattering angle is always $\theta_{xy}=90^\circ$ and
the surface brightness increases for higher
inclinations as in the inclined surface emissivity $\propto 1/\cos i$. 

The scattering asymmetry parameter
$g$ is relatively small and therefore the front--back brightness
differences are not strong in Fig.~\ref{FigDisk2Dg3}. The forward
scattering effect is much clearer in Fig.~\ref{FigDisk2Di45}, where
two disks are plotted with the same $i$ and $\epsilon(r)$, but for
isotropic scattering $g=0$ and strong forward scattering $g=0.6$. 

For the disks shown in
Figs.~\ref{FigDisk2Dg3} and \ref{FigDisk2Di45}, the absolute signal drops
in the radial direction for all azimuthal angles $\phi_{xy}={\rm atan2}(x,y)$ by
exactly a factor of two from the inner edge $(r_1)_{xy}$ 
to the outer edge $(r_2)_{xy}$ according to the adopted radial dependence of the
scattering emissivity $\epsilon(r)\propto 1/r$. This is equivalent to
the statement that the (relative) azimuthal dependence along the
ellipse $r_{xy}$ describing a ring in the inclined disk is the same
for all separations in a given disk image $I(r_{xy},\varphi_d)$,
$Q_\phi(r_{xy},\varphi_d)$, $Q_{\rm d}(r_{xy},\varphi_d)$, or $U_{\rm d}(r_{xy},\varphi_d)$. 
This is also valid if the azimuthal angle $\varphi_d$ for the disk
plane is replaced by the on-sky azimuthal angle $\phi$.

Therefore, it is possible to determine the azimuthal dependence of the
scattered light in a very simple way using azimuthal phase functions
defined in the disk plane, without considering the radial distribution
of the scattering emissivity. 

\subsubsection{Scattering phase functions for the disk azimuth angle}

The azimuthal dependence of the scattered light can be
calculated easily for flat, rotationally symmetric, and optically
thin disks as a function of the azimuthal angle $\varphi_d$ in the disk plane.
For this, we have to express the scattering angle $\theta$ 
as a function of $\varphi_d$ and the disk inclination $i$ according to
\begin{equation}
  \theta_{\varphi,i}=\theta(\varphi_d,i) = {\rm acos}
                     (\cos \varphi_d \cdot (-\sin i))\,,
\label{Eqthetaphi}
\end{equation}
where $\varphi_d=0$ for the far-side semi-minor axis of the
projected disk.  

\begin{figure}
\includegraphics[trim=2cm 13cm 8cm 2cm, width=8.8cm]{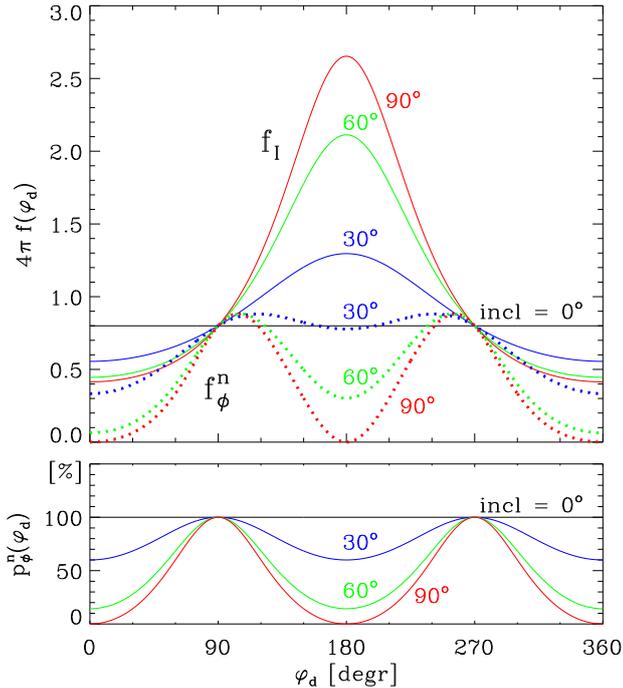}
\caption{Upper panel: Azimuthal dependence of the disk scattering phase
  function for the intensity $4\pi\,f_I(\varphi_d,i,g)$ and the
  normalized ($p_{\rm max}=1$) polarized intensity $4\pi\,f_\phi^n(\varphi_d,i,g)$
  for disks with scattering asymmetry parameter $g=0.3$ and 
  inclinations $i=0^\circ,~30^\circ,~60^\circ,$ and $90^\circ$. Lower panel:
  Normalized fractional polarization $p^n_\phi(\varphi_d,i) = f^n_\phi/f_I$
  for the same inclinations; $p^n_\phi$ does not depend on the $g$-parameters.}
\label{Fig1Ring2D}
\end{figure}

The dependence of the scattered intensity and
polarization flux with disk azimuthal angle $\varphi_d$ follows directly
from the scattering phase functions $f_I$ and $f_\phi$ and
the change of the scattering angle $\theta$ as a function of $\varphi_d$ and $i$
\begin{eqnarray}
&&\hspace{-0.7cm} f_I(\varphi_d,i,g)= f_I(\theta_{\varphi,i},g) \,,\\
&&\hspace{-0.7cm} f_\phi(\varphi_d,i,g,p_{\rm max})=p_{\rm max}\,f^n_\phi(\theta_{\varphi,i},g)= f_\phi(\theta_{\varphi,i},g,p_{\rm max}) \,.
\end{eqnarray}  

Figure~\ref{Fig1Ring2D} shows the azimuthal scattering
functions $4\pi\,f_I(\varphi_d,i,g)$
for the intensity and $4\pi\,f_\phi^n(\varphi_d,i,g)$ for
the polarized flux for $g=0.3$ and for different inclinations.
The factor $4\pi$ normalizes the
isotropic scattering case $4\pi\,f_I(\varphi_d,i,$g=0$)=1$ and scales all
other cases $g\neq 0$ accordingly. 

The enhanced forward scattering
for $f_I(\varphi_d)$ around $\varphi_d=180^\circ$ is clearly visible
for inclined disks.
The polarization function $f^n_\phi(\varphi_d)$ has for
backward scattering around $\varphi_d=0^\circ$ and
  forward scattering around $\varphi_d=180^\circ$ strongly reduced
  values in highly inclined disks
when compared to the intensity as can be seen for the green and red
curves in Fig.~\ref{Fig1Ring2D}.
At $\varphi_d=90^\circ$ and $270^\circ$, the functions $f_I$ and $f^n_\phi$
have (for given $g$) the same value for all inclinations
because the scattering angles are always $\theta_{\varphi,i}=90^\circ$
($\cos \varphi_d=0$ in Eq.~\ref{Eqthetaphi}).

The azimuthal dependence of the fractional polarization is
given by
\begin{equation}
  p^n_\phi(\varphi_d,i) = \frac{p_{\rm sca}(\theta_{\varphi,i})}{p_{\rm max}}
         =\frac{f^n_\phi(\varphi_d,i,g)}{f_I(\varphi_d,i,g)}
,\end{equation}  
and this function does not depend on the asymmetry parameter $g$. This
dependence is equivalent to Rayleigh scattering and is shown
in Fig.~\ref{Fig1Ring2D} for completeness.

For the phase function $f_Q$ and $f_U$ one needs to consider that
the Stokes $Q_{\rm d}$ and $U_{\rm d}$ parameters are defined in the disk
coordinate system projected onto the sky while $f_\phi$
is given for the azimuthal angle $\varphi_d$ for the
$x_d,y_d$-coordinates of the disk midplane.
For the splitting of $f_\phi(\varphi_d,i,g,p_{\rm max})$
into $f_Q$ and $f_U$, the azimuthal angle $\phi_{xy}$
defined in the $x$-$y$ sky plane must be used.
The relation between $\varphi_d$ and $\phi_{xy}$ is
\begin{equation}
  \phi_{xy}(\varphi_d,i)={\rm atan}\Biggl(\frac{\tan\varphi_d}{\cos i}\Biggr)\,.
\label{Eqphiphi}
\end{equation}

The phase functions for $Q_{\rm d}$ and $U_{\rm d}$ are then equivalent to
the conversion given in Eqs.~\ref{EqQQphi} and \ref{EqUQphi}: 
\begin{eqnarray}
f_Q(\varphi_d,i,g,p_{\rm max})
       =-f_\phi(\varphi_d,i,g,p_{\rm max}) \cdot \cos(2\,\phi_{xy}(\varphi_d,i)) \\
f_U(\varphi_d,i,g,p_{\rm max})
       =-f_\phi(\varphi_d,i,g,p_{\rm max}) \cdot \sin(2\,\phi_{xy}(\varphi_d,i)) \,.
\end{eqnarray}

\begin{figure}
  \includegraphics[trim=2cm 13cm 8cm 2cm, width=8.8cm]{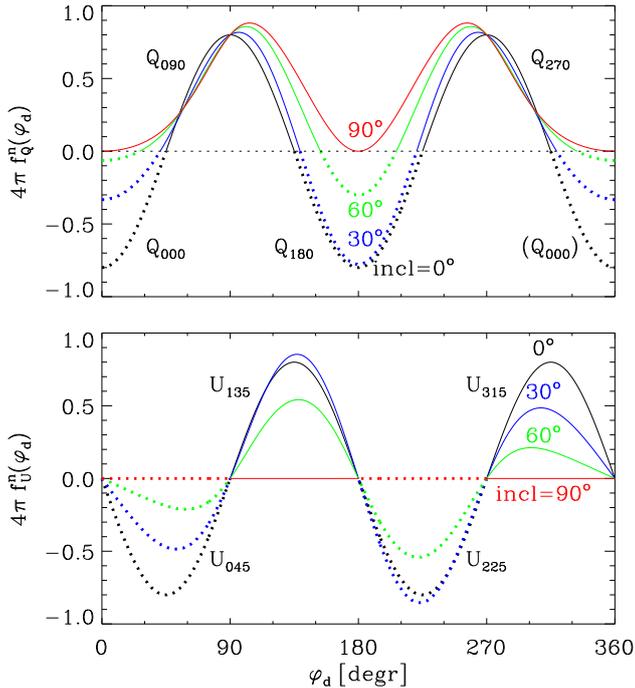}
  \caption{Azimuthal dependence for the Stokes scattering phase functions
  $4\pi \cdot f^n_Q(\varphi_d,i,g)$ and 
  $4\pi \cdot f^n_U(\varphi_d,i,g)$ for a rotationally symmetric,
  flat, optically thin disk with scattering asymmetry parameter $g=0.3$
  and inclinations $i=0^\circ,~30^\circ,~60^\circ,$ and $90^\circ$.
  }
\label{Fig2Ring2D}
\end{figure}

These azimuthal function of the ``on-sky''
Stokes parameters $f_Q$ and $f_U$ is shown in Fig.~\ref{Fig2Ring2D}.
The functions are characterized by their double wave,
which for $i=0^\circ$  are exact double-wave cosine 
$f_Q(\varphi_d)\propto -\cos 2\varphi_d$ and double-wave sine
$f_U(\varphi_d)\propto -\sin 2\varphi_d$ functions. Deviations from the
sine and cosine function become larger with increasing $i$, particularly
for large asymmetry parameters $g$. The positive and negative
sections of the $f_Q(\varphi_d)$ and $f_U(\varphi_d)$ functions
correspond to the positive and negative polarimetric quadrants.
The $f_Q$ and $f_U$ functions can also be expressed as normalized
functions $f^n_Q$ and $f^n_U$ for $p_{\rm max}=1$ and as fractional
polarization $p^n_Q$ and $p^n_U$ equivalent to $p^n_\phi$ given above.

For the adopted HG$_{\rm pol}$
dust scattering phase function, these phase functions $f_{\varphi}$ provide
a universal description of the azimuthal flux and polarized flux
dependence as a function of inclination $i$
for all flat, optically thin, rotationally symmetric disks. 

\subsubsection{Disk-averaged scattering functions}

The total intensity $\overline{I}$ for our disk model
can be conveniently calculated in $r,\varphi_d$-coordinates because
the integration can be separated between the $r$-dependent
scattering emissivity $\epsilon$ and the $\varphi_d$-dependent
scattering phase function $f_I$ according to
\begin{equation}
  \overline{I}
  = 2\pi \int_{0}^{\infty} \epsilon(r)\, r\,{\rm d}r 
  \cdot \frac{1}{2\pi} \int_0^{2\pi} f_I(\varphi_d,i,g)\, {\rm d}\varphi
  = \overline{\epsilon} \cdot \langle f_I(i,g) \rangle  \,.
 \label{Eqintphir}
\end{equation}                
The first term represents the total
scattering emissivity $\overline{\epsilon}$ of the disk, and
the second term is the disk averaged scattering phase function
for the intensity $\langle f_I(i,g) \rangle$.

For the integrated polarization parameters, the same type of relation
can be used
\begin{eqnarray}
  \overline{Q}_\phi=\overline{\epsilon} \cdot
  \langle f_\phi(i,g,p_{\rm max}) \rangle \,, \label{Eqintqphi}\\
  \overline{Q}_{\rm d}=\overline{\epsilon} \cdot
  \langle f_Q(i,g,p_{\rm max}) \rangle \,, \label{Eqintq}\\
  \overline{U}_{\rm d}=\overline{\epsilon} \cdot
  \langle f_U(i,g,p_{\rm max}) \rangle \,.
\end{eqnarray}  
The scale factor $p_{\rm max}$
can be separated from the normalized versions of the
disk-averaged scattering functions as
\begin{equation}
\langle f_\phi(i,g,p_{\rm max}) \rangle = p_{\rm max}\, \langle f^n_\phi(i,g) \rangle,
\end{equation}
and similar for $f_Q$ and $f_U$.
The disk-averaged fractional polarization follows from
\begin{equation}
  \langle p_\phi(i,g,p_{\rm max}) \rangle =
          p_{\rm max}\, \langle p^n_\phi(i,g) \rangle=
  \frac{\langle f_\phi(i,g,p_{\rm max}) \rangle}{\langle f_I(i,g) \rangle}\,,
\end{equation}
and similar for $p_Q$ or $p_U$, where the latter is always zero
for rotationally symmetric disks. Unlike for the azimuthal dependence
of the fractional polarization $p^n(\varphi_d,g),$ the disk-averaged parameters
$\langle p^n_\phi(i,g) \rangle$ depend on the $g$-parameter because  in this average
$g$ shifts the flux weight between disk regions producing
higher or lower levels of scattering polarization. 

\subsection{Normalized quadrant polarization parameters}
\label{NormQuad1}
The Stokes $Q_{xxx}$ and $U_{xxx}$ quadrant polarization parameters
correspond to the individual positive and negative sections of
the $f_Q(\varphi_d)$ and $f_U(\varphi_d)$ disk phase function shown
in Fig.~\ref{Fig2Ring2D}. The relation between quadrant
parameters and phase function follow the same scheme
as for the disk-integrated quantities $\overline{Q}_{\rm d}$
and $\overline{U}_{\rm d}$ described by Eq.~\ref{Eqintphir}
but the integration is limited to the azimuthal angle range
$\varphi_1$ to $\varphi_2$ of a given quadrant instead of
0 to $2\pi$. For Stokes $Q_{\rm d}$, there is
\begin{equation}
  Q_{xxx}
  = (\varphi_2-\varphi_1) \int_{0}^{\infty} \epsilon(r)\, r\,{\rm d}r 
  \cdot \frac{p_{\rm max}}{\varphi_2-\varphi_1}
             \int_{\varphi 1}^{\varphi 2} f_Q^n(\varphi_d,i,g)\, {\rm d}\varphi\,.
 \label{Eqintqnorm}
\end{equation}    
The first term is the disk scattering emissivity $\epsilon$
integrated for the quadrant and the second term is the averaged
$f_Q^n$ scattering phase function for this quadrant. Because $\epsilon$
is independent of the azimuthal angle, the first term can be expressed
as a fraction of the disk integrated emissivity
$\overline{\epsilon}\cdot(\varphi_2-\varphi_1)/2\pi$ and the
equation takes the form 
\begin{equation}
  Q_{xxx}=\overline{\epsilon}\,p_{\rm max}\, \frac{1}{2\pi}
  \int_{\varphi 1}^{\varphi 2} f_Q^n(\varphi_d,i,g)\, {\rm d}\varphi
                   =\overline{\epsilon}\, p_{\rm max}\,Q_{xxx}^n(i,g)\,,
\label{EqNormQuad}
\end{equation}  
where we introduce the normalized quadrant polarization $Q_{xxx}^n$.
The same scaling factors $\overline{\epsilon}$ and $p_{\max}$
are involved as for the equations for the integrated polarization parameters
$\overline{Q}_\phi$, $\overline{Q}_{\rm d}$, and $\overline{U}_{\rm d}$, and
therefore $Q^n_{xxx}$ and $\langle f_Q(i,g) \rangle$ are related
with the same factors $\overline{\epsilon}$ and $p_{\rm max}$
to observed quantities $Q_{xxx}$ and $\overline{Q}_{\rm d}$.
The same formalism applies to the Stokes $U_{\rm d}$ quadrants.

According to Eq.~\ref{EqNormQuad}, the normalized polarization quadrants
$Q^n_{000}$, $Q^n_{090}$, $Q^n_{180}$, $Q^n_{270}$ and
$U^n_{045}$, $U^n_{135}$, $U^n_{225}$, $U^n_{315}$ depend only on $i$ and $g$
as in the disk-averaged functions $\langle f^n_Q \rangle$
or $\langle f^n_U\rangle$. The azimuthal integration range
$\varphi_1$ to $\varphi_2$ is of course different for each
quadrant, as summarized in Table~\ref{Integration}. For the
$Q_{\rm d}$-quadrants, the geometric projection
effect introduces an inclination dependence for the
integration boundaries $\varphi_1$ and $\varphi_2$. For
increased $i,$ the fraction of the sampled disk surfaces
$s_{xxx}=(\varphi_1(i)-\varphi_2(i))/2\pi$ in the
left and right quadrants $Q^n_{090}$ and $Q^n_{270}$ is
enhanced $s_{090}=s_{270}=0.5-{\rm atan}(\cos i)/\pi$
and the disk surface fraction located in the front and back
quadrants $Q^n_{180}$ and $Q^n_{000}$ is reduced
$s_{000}=s_{180}={\rm atan}(\cos i)/\pi$, respectively (we note that
atan$(\cos 0^\circ)=\pi/4$ and atan$(\cos 90^\circ)=0$).
There is no $i$-dependence for the splitting of the Stokes $U_{\rm d}$ quadrants,
because the ``left-side'' back and front quadrants and
the ``right-side'' back and front quadrants sample the same disk
surface fraction $s_{045}=s_{135}=s_{225}=s_{315}=0.25$ for
all inclinations.

\begin{table}
  \caption{Integration ranges for Stokes $Q_{\rm d}$ and $U_{\rm d}$ polarization
    quadrants for on-sky azimuthal angles $\phi_{x,y}={\rm atan2}(y,x)$ and disk
    azimuthal angles $\varphi_d={\rm atan2}(y_d,x_d)$.}
\label{Integration}
\begin{tabular}{lcccc}
\hline\hline
\noalign{\smallskip}
    &  \multispan{2}{\hfil \hspace{-0.2cm}$\int_{\phi1}^{\phi2}Q_{\rm d}\,{\rm or}\, U_{\rm d}(\phi_{xy})d\phi$\hspace{-0.2cm}\hfil}
    & \multispan{2}{\hfil $\int_{\varphi1}^{\varphi2}f_{Q\,{\rm or}\,U}(\varphi_d)d\varphi$\hfil} \\
  &  $\phi1$  & $\phi2$  & $\varphi1$  & $\varphi2$  \\ 
\noalign{\smallskip\hrule\hrule\smallskip}
$Q_{000}$ & $-45^\circ$ & $ 45^\circ$ & $-{\rm atan}(\cos i)$
                                        &  ${\rm atan}(\cos i)$ \\
$Q_{090}$ & $45^\circ$  & $135^\circ$ &  ${\rm atan}(\cos i)$
                                        &  $180^\circ$--${\rm atan}(\cos i)$  \\
$Q_{180}$ & $135^\circ$ & $225^\circ$ & $180^\circ$--${\rm atan}(\cos i)$
                                        &  $180^\circ$+${\rm atan}(\cos i)$  \\
$Q_{270}$ & $225^\circ$ & $315^\circ$ & $180^\circ$+${\rm atan}(\cos i)$
                                        &  $360^\circ$--${\rm atan}(\cos i)$ \\
\noalign{\smallskip}
$U_{045}$ & $ 0^\circ$  & $90^\circ$  & $0^\circ$    & $90^\circ$            \\
$U_{135}$ & $90^\circ$  & $180^\circ$ &  $90^\circ$  & $180^\circ$            \\
$U_{225}$ & $180^\circ$ & $270^\circ$ &  $180^\circ$ & $270^\circ$            \\
$U_{315}$ & $270^\circ$ & $360^\circ$ &  $270^\circ$ & $360^\circ$           \\
\noalign{\smallskip\hrule\hrule\smallskip}
\end{tabular}
\end{table}

\section{Calculation of disk polarization parameters}

The previous section shows that the disk-integrated
radiation parameters such as $\overline{I},\overline{Q}_\phi$
and the quadrant polarization parameters
$Q_{xxx}$ and $U_{xxx}$ can be expressed as 
disk-averaged phase functions $\langle f(i,g) \rangle$ and
normalized polarization parameters $Q_{xxx}^n(i,g)$ and
$U_{xxx}^n(i,g)$ describing the azimuthal dependence
of the scattering, and scale factors for the total
disk-scattering emissivity $\overline{\epsilon}$ and the
maximum scattering polarization $p_{\rm max}$. 

This simplicity allows a concise but comprehensive
graphical presentation covering the full parameter
space of the disk model parameters for the
intensity and polarization, including the newly introduced
quadrant polarization parameters. The Appendix gives a few 
IDL code lines for calculation of the
numerical values. In addition, we explore the deviation of the
results from vertically extended disk models and from the flat disk models.

\begin{figure}
\includegraphics[trim=3.0cm 12.5cm 8.5cm 3.5cm, width=8.8cm]{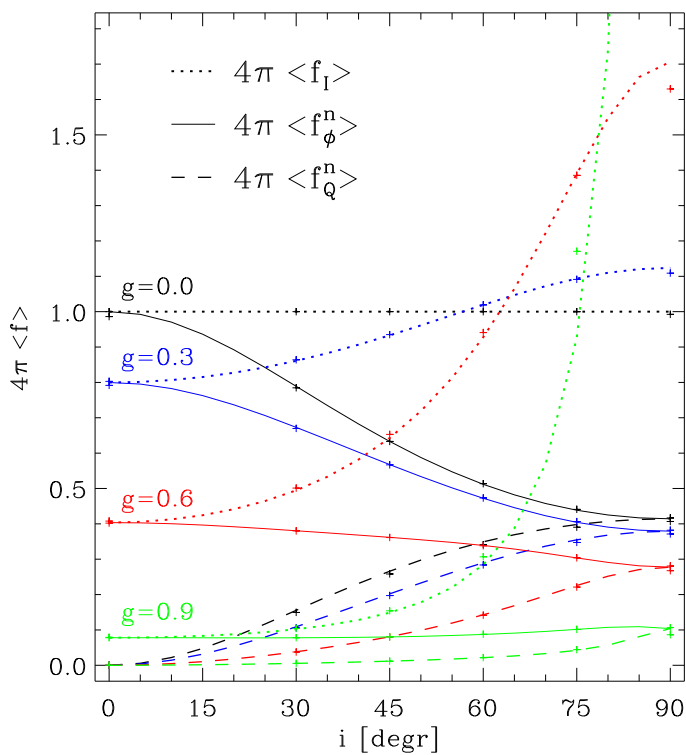}
\caption{Disk-averaged scattering phase functions for the intensity
  $4\pi\,\langle f_I \rangle$, the normalized ($p_{\rm max}$=1)
  azimuthal polarization $4\pi\,\langle f_\phi^n \rangle$,
  and the normalized Stokes $Q_{\rm d}$ flux $4\pi\,\langle f_Q^n \rangle$
  versus the disk inclination for HG-asymmetry parameters
  $g=0$ (black), 0.3 (blue), 0.6 (red), and 0.9 (green). Lines
  give the results for flat disks and crosses for vertically
  extended disk rings (Sect.~\ref{Sect3d}, Fig.~\ref{FigModelsg6}).}
\label{3ddisk_FigApr1}
\end{figure}

\subsection{Calculations of the disk-averaged intensity and
  polarization scattering functions}
\label{Sectap}
The disk-averaged scattering phase functions $\langle f(i,g) \rangle$
are equivalent to the basic integrated quantities
$\overline{I}$, $\overline{Q}_\phi$, and $\overline{Q}_{\rm d}$ if
normalized scale factors $\overline{\epsilon}=1$ and $p_{\rm max}=1$
are used  (Eqs.~\ref{Eqintphir} to \ref{Eqintq}). 
The functions $\langle f_I(i,g) \rangle$,
$\langle f^n_\phi(i,g) \rangle,$ and $\langle f_Q^n(i,g) \rangle$
are plotted in Fig.~\ref{3ddisk_FigApr1} as a function of inclination
for the four asymmetry parameters $g=0,\,0.3,\,0.6,$ and 0.9.
The plot includes the results 
from calculation of vertically extended, three-dimensional disk rings (crosses)  for comparison,
which are discussed in Sect.~\ref{Sect3d}. The results
for $\langle f(i,g) \rangle$
are multiplied by the factor $4\pi$ because this sets the special
reference case for isotropically scattering dust $g=0$ for all
inclinations to
$4\pi \langle f_I(i,0) \rangle=4\pi \langle f^n_\phi(i,0) \rangle=1$
and simplifies the discussion.

For pole-on disks $i=0^\circ$, there is
$\langle f^n_\phi(0^\circ,g) \rangle = \langle f_I(0^\circ,g) \rangle $
 for all $g$ parameters because the scattering angle is $\theta=90^\circ$  everywhere.
For enhanced scattering asymmetry parameters $g>0$ (but also for
$g<0$), the disk intensity is below the isotropic case
$4\pi\langle f_I \rangle<1$ for lesser and moderately inclined disks
$i\lapprox 60^\circ$, because the enhanced forward scattering
(or backward scattering) produces enhanced scattered flux in
directions near to the disk plane and reduced flux for polar viewing
angles. 

The forward scattering enhances the scattered intensity to
$4\pi \langle f_I \rangle >1$ for $i\ga 60^\circ$
and this effect becomes particularly strong for
$g\rightarrow 1$ and $i\rightarrow 90^\circ$.
This behavior is well known and produces a strong detection bias
for high-inclination debris disks
\citep[e.g.,][]{Artymowicz89,Kalas96,Esposito20}.

For the polarized flux $\langle f^n_\phi(i,g) \rangle,$ an
enhanced inclination does not produce an enhancement
of the $\langle f_\phi\rangle$ signal because the strong increase
in scattered flux from the forward scattering direction (or backward
direction for $g<1$) is predominantly unpolarized. For moderate asymmetry parameter $|g|\lapprox 0.6,$ this causes
an overall
decrease of $\langle f_\phi\rangle$ with inclination
(Fig.~\ref{3ddisk_FigApr1})
while for extreme values $|g|\gapprox 0.9$ the huge
flux increase compensates for the lower fractional polarization for
forward and backward scattering.

The phase function for the Stokes $Q_{\rm d}$ parameter
$\langle f^n_Q(0^\circ,g) \rangle$ is zero for the pole-on view because
of the symmetric cancellation of $+Q_{\rm d}$ and $-Q_{\rm d}$ signals. The
function increases steadily with $i$ (Fig.~\ref{3ddisk_FigApr1})
and for $i=90^\circ$ or edge-on disks there is
$\langle f^n_Q(90^\circ,g) \rangle = \langle f^n_\phi(90^\circ,g) \rangle$,
because all dust is aligned with the major axis and
produces polarization in the $+Q_{\rm d}$ direction.

\begin{figure}
\includegraphics[trim=2.0cm 13cm 8cm 3.5cm, width=8.8cm]{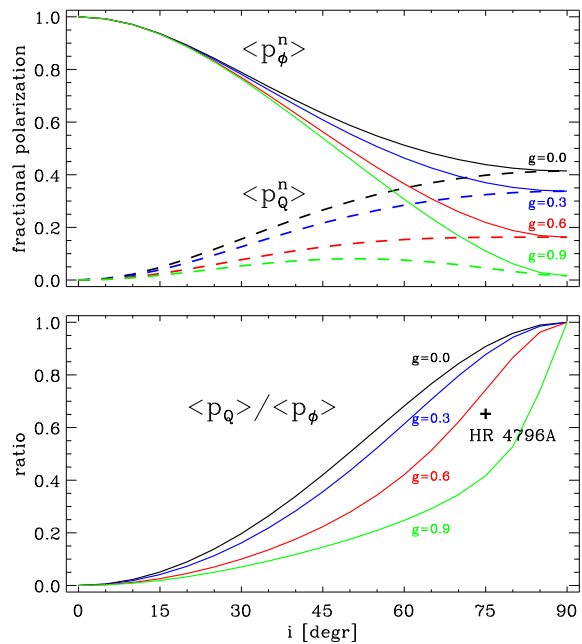}
\caption{Upper panel: Disk averages of the fractional
  azimuthal polarization
  $\langle p_\phi^n\rangle$ and the Stokes $Q_{\rm d}$ polarization
  $\langle p_Q^n\rangle$
  normalized for $p_{\rm max}=1$ for different asymmetry parameters and
  as function of inclination. Lower panel: Polarization ratios 
  $\langle p_Q\rangle/\langle p_\phi\rangle=\overline{Q}_{\rm d}/\overline{Q}_\phi$
  and the corresponding measurement for HR~4796A.}
\label{3ddisk_FigApr2}
\end{figure}

\paragraph{Fractional polarization.} The fractional polarizations
$\langle p_{\phi}^n(i,g)\rangle$ and $\langle p_Q^n(i,g)\rangle$
in Fig.~\ref{3ddisk_FigApr2} can be deduced from the ratio of the
phase functions shown in Fig.~\ref{3ddisk_FigApr1}. Observationally,
a fractional polarization determination requires a measurement of
the integrated disk polarization and the integrated disk intensity. 

As shown in Fig.~\ref{3ddisk_FigApr2}, the fractional azimuthal
polarization $\langle p_{\phi}^n(i,g)\rangle$  only depends to a very small extent on $g$
for small inclinations $i\le 35^\circ$ with deviations $<\pm 0.02$
and the $i$-dependence is well described by
\begin{equation}
\frac{\overline{Q}_\phi}{\overline{I}}=p_{\rm max} \langle p_{\phi}^n(i)\rangle \approx p_{\rm max} \cdot\, (\cos i)^{1.7}\,.
\label{pincl}
\end{equation}
A measurement of the fractional azimuthal polarization for low-inclination disks  is therefore equivalent to a determination of the
$p_{\rm max}$-parameter.

The lower
panel of Fig.~\ref{3ddisk_FigApr2} includes also the
purely polarimetric ratio 
$\langle p_Q\rangle / \langle p_{\phi}\rangle =
\overline{Q}_{\rm d}/\overline{Q}_\phi$,
which includes no scaling factor $p_{\rm max}$ and systematic uncertainties
from the polarimetric measurements might be particularly small.
Therefore, the ratio
$\overline{Q}_{\rm d}/\overline{Q}_\phi$ can be used to determine
the scattering asymmetry parameter $g$, if polarimetric cancellation effects
for $\overline{Q}_\phi$ are taken into account for poorly
resolved disks. For the extended disk HR~4796A, cancellation can be
neglected, and we can use $\overline{Q}_{\rm d}/\overline{Q}_\phi=0.652\pm 0.030$
($\Sigma Q_{xxx}/\overline{Q}_\phi$ from Tab.~\ref{QuadHR4796A}) and the
inclination $i=75^\circ$, which yields a value of about $g=0.7$
from Fig.~\ref{3ddisk_FigApr2}. This method is useful 
for systems with $i\approx 30^\circ-80^\circ$ because the separations
between the $g$-parameter curves are quite large. The curves in Fig.~\ref{3ddisk_FigApr2} 
for flat disk models are not applicable for edge-on disks $i\ga 80^\circ$ with a
vertical extension (see Sect.~\ref{Sect3d}).

\begin{figure*}
\includegraphics[trim=2.2cm 13cm 2cm 3.5cm, width=15cm]{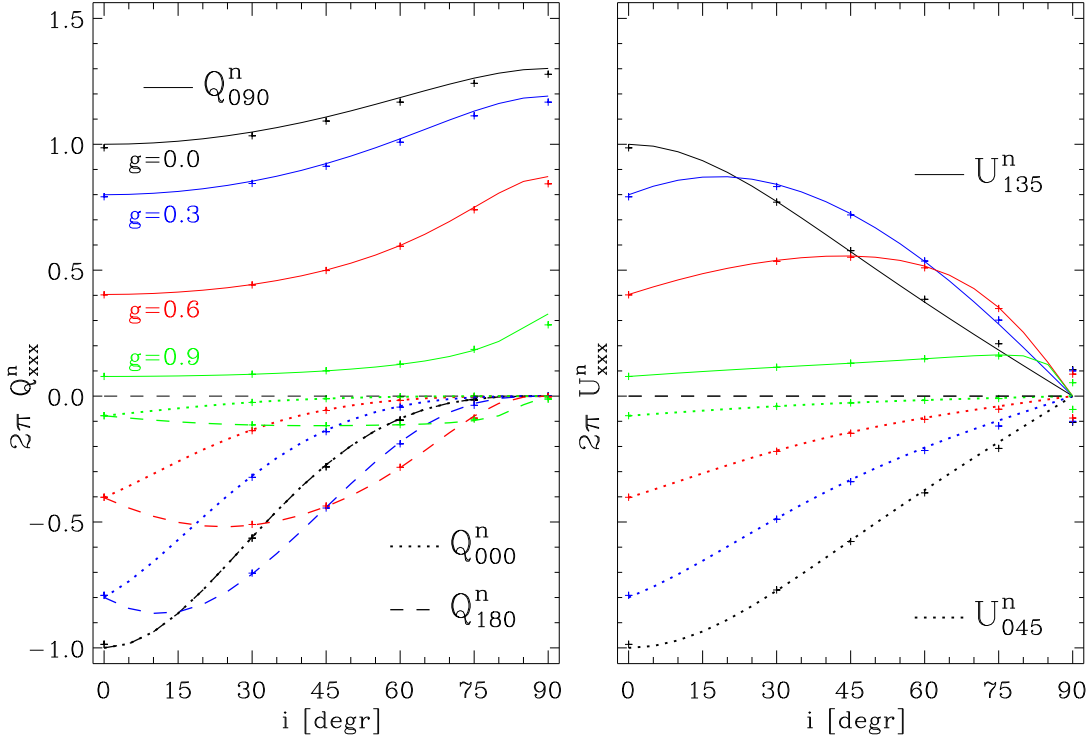}
\caption{Normalized ($p_{\rm max}=1,\, \overline{\epsilon}=1$)
  quadrant polarization parameters for $2\pi\,Q^n_{000}$,
  $2\pi\,Q^n_{090}$, $2\pi\,Q^n_{180}$ (left)
  and $2\pi\,U^n_{045}$, $2\pi\,U^n_{135}$ (right) 
  as function of disk inclination and for the HG-asymmetry parameter
  $g=0$ (black), 0.3 (blue), 0.6 (red), and 0.9 (green). Lines give
  the results for flat disks and crosses for vertically extended disk
  rings (Sect.~\ref{Sect3d}, Fig.~\ref{FigModelsg6})}  
\label{FigQuadrants1}
\end{figure*}

\subsection{Calculations of the normalized quadrant polarization parameters}
\label{NormQuad2}
The normalized quadrant polarization parameters
$Q^n_{000}$, $Q^n_{090}$, $Q^n_{180}$, and $U^n_{045}$ $U^n_{135}$ 
are plotted in Fig.~\ref{FigQuadrants1} as a function of
$i$ for different $g$ parameters.
The ``right-side'' quadrants $Q^n_{270}$, $U^n_{315}$, $U^n_{225}$
have the same absolute values as
the corresponding ``left-side'' quadrants because of the disk symmetry. 
All values are multiplied by the factor $2\pi$
so that the reference case $g=0$ and $i=0^\circ$ is set to 
$ 2\pi\, |Q^n_{xxx}(0^\circ,0)|= 2\pi\, |U^n_{xxx}(0^\circ,0)|=1,$ similar
to the normalization for the disk-averaged
scattering function $4\pi\,\langle f_\phi^n(i,0) \rangle$.

For pole-on disks, all the normalized
quadrant polarization values have the same absolute value $|Q^n_{xxx}(0^\circ,g)|=|U^n_{xxx}(0^\circ,g)|$  for a given $g$
.
This value is lower for larger $g$-parameter
because less light is scattered perpendicularly to
the disk plane in the polar direction, as in the
$\langle f^n_\phi\rangle$ function. For $i> 0^\circ$ 
the quadrant values show different types of
dependencies on $i$ and $g$. 

As described in Sect.~\ref{NormQuad1}, for the Stokes $Q_{\rm d}$ quadrants the disk inclination introduces
 a geometric projection effect
which increases the sampled disk area
for the ``left'' and ``right'' quadrants  for larger $i$ , and therefore the
$Q^n_{090}$ and $Q^n_{270}$-values reach a maximum of ($Q_\phi^n/2$)  for $i=90^\circ$
. The areas for the front and
back quadrants go to zero for $i\rightarrow 90^\circ$ and therefore
so do the values $Q^n_{180}$ and $Q^n_{000}$, but with a difference
which depends strongly on $g$. The effect of the scattering
asymmetry is already clearly visible for relatively small values,
$g\approx 0.3,$ and inclinations, $i\approx 10^\circ$, and becomes even
stronger for larger $g$ and $i$ as can be seen from the enhanced
$|Q_{180}|$ values relative to $|Q_{000}|$. A similar front--back
quadrant effect also occurs for the $U_{\rm d}$-components with enhanced
absolute values for the front-side quadrant
$|U_{135}|$ and reduced values for the backside
quadrant $|U_{045}|$.   

\begin{figure*}
  \includegraphics[width=8.8cm]{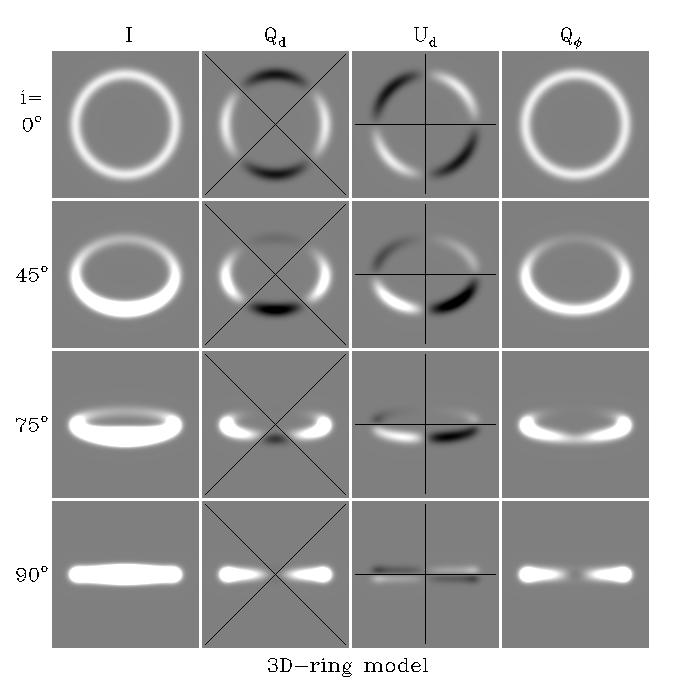}
  \includegraphics[width=8.8cm]{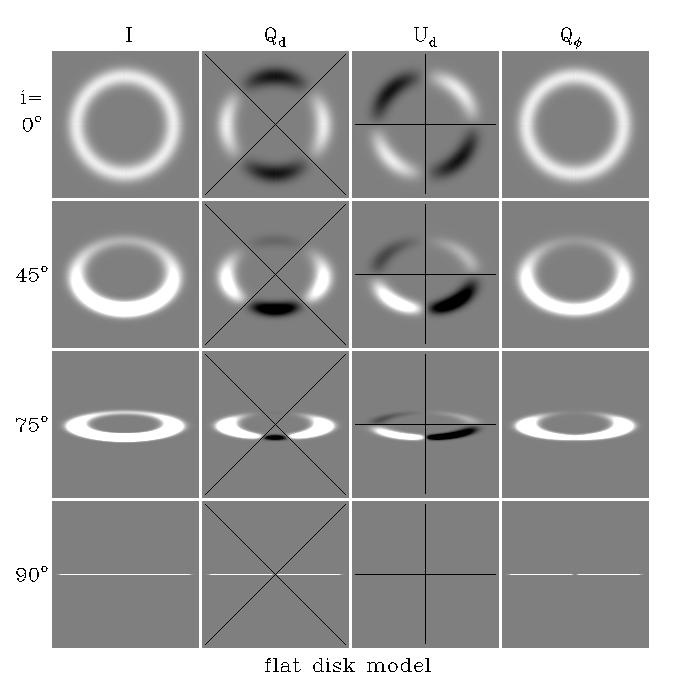}
  \caption{Comparison of the 3D-ring model with the flat disk model.
  Left: $I$, $Q_{\rm d}$, $U_{\rm d}$ and $Q_\phi$ for a 3D disk
  ring with a full width at half maximum density distribution of
  $\Delta_{\rm FWHM}/r_{\rm ring}=0.2$. Scattering parameters are $g=0.6$
  and $p_{\rm max}=1$ for inclinations $i=0^\circ$, $45^\circ$, $75^\circ$
  and $90^\circ$. The same gray scale from $+a$ (white) to $-a$
  (black) is used for all panels. The black lines in the $Q_{\rm d}$ and $U_{\rm d}$
  panels indicate the polarization quadrants. Right: Same but for
  the flat disk model.}
\label{FigModelsg6}
\end{figure*}

\subsection{Comparison with three-dimensional disk ring models}
\label{Sect3d}
The polarization parameters derived in the previous sections are
calculated for geometrically flat disks because this simplifies
the calculations enormously. Of course, real disks have 
a vertical extension but observations of highly inclined debris
disks typically show a small ratio $h/r\la 0.1$
\citep[see e.g.,][]{Thebault09}. Therefore, the flat disk models could serve
as an approximation for 3D disks, and
we explore the differences. For this, we calculated models for a
rotationally symmetric, optically thin
disk ring with a central radius $r_{\rm ring}$
and a Gaussian density distribution for the ring cross-section,
\begin{equation}
  \rho(r,h) = \rho_0\, {\rm exp}(-[(r-r_{\rm ring})^2+h^2]/2\,\delta^2) \,,
\end{equation}
with full width at half maximum (FWHM) of
$\Delta_{\rm FWHM}=2.355\,\delta$. Figure~\ref{FigModelsg6} compares
images of such 3D disk rings with $\Delta_{\rm FWHM}=0.2\,r_{\rm ring}$,
$g=0.6,$ and different $i$ with flat disk models. The vertical extension
of the 3D model is most obvious for the edge-on ($i=90^\circ$) case
for which the flat disk model gives only a profile along
the $x$-axis.

For $i<90^\circ$, the differences between flat disks and vertically
extended disks are already small for $i=75^\circ$ and hardly recognizable for
lower inclination. In particular, the values for the disk-averaged scattering
functions $\langle f \rangle$ and the normalized quadrant polarization
parameters $Q_{xxx}^n$ and $U_{xxx}^n$ are equal or very similar
as can be seen in Figs.~\ref{3ddisk_FigApr1} and \ref{FigQuadrants1}
where the results from the 3D disks are plotted as small
crosses together with those from the flat disk models. 
The agreement is typically better than $\pm 0.01$. An example of a
systematic difference between 3D disks and flat disks is
a slightly lower value (about 0.01) for the azimuthal polarization
$\langle f_\phi \rangle$ in Fig.~\ref{3ddisk_FigApr1} for pole-on
($i=0^\circ$) 3D disks. For disks with a vertical extension, not all scatterings
are occurring exactly in the disk midplane, but also slightly above
and below where the scattering angle is smaller or larger than
$90^\circ$ and therefore $(1-\cos^2\theta)/(1+\cos^2\theta)$ in
Eq.~\ref{Eqpsca} is smaller than one. Another example is the reduced
$\langle f_I \rangle$ for edge-on ($i=90^\circ$) 3D disks,
because less material lies exactly in front of
the star where forward scattering would produce a strong maximum for
large $g$-parameter. This is most visible for $g=0.6$ (the effect is even stronger for $g=0.9$
 but this point is outside the plotted range).
Similar but typically also very
small effects are visible for the normalized quadrant polarization
parameters in Fig.~\ref{FigQuadrants1}. The presence or absence
of the vertical extension produces a strong difference for the Stokes $U_{\rm d}$
which is also clearly apparent in Fig.~\ref{FigModelsg6} for $i=90^\circ$.

These comparisons show that the quadrant polarization
parameters derived from flat, optically thin, rotationally symmetric
disk models are for most cases essentially indistinguishable
from those derived from
3D disk models with a vertical extension typical for debris disks. Only
for edge-on or close to edge-on disks, $i\ga 80^\circ$, can  the vertical extension
introduce significant differences, which needs to
be taken into account.

\section{Diagnostic diagrams for the scattering asymmetry $g$}

The normalized quadrant polarization parameters and quadrant
ratios for disk models using the HG$_{\rm pol}$-function
depend only on the scattering asymmetry
parameter $g$ and the disk inclination $i$
as shown in Fig.~\ref{FigQuadrants1}. From
imaging polarimetry of debris disks one can often
accurately measure the disk inclination, and therefore the quadrant polarization
parameters are ideal for the determination of $g$. The method described
in this work for the single parameter HG$_{\rm pol}$-function
can be generalized to other, more sophisticated parameterizations
for the polarized scattering phase function of the dust.

This diagnostic method is based on the strong assumption that
the intrinsic disk geometry is rotationally symmetric,
which is often a relatively good assumption for debris disks, 
but there are also several cases known with significant
deviations from axisymmetry \citep[e.g.,][]{Debes09,Maness09,Hughes18}.
This can affect the determination of the $g$-parameter and one
should always assess possible asymmetries in the disk symmetry
using for example the left--right symmetry parameters
$\Delta^{090}_{270}$ or $\Delta^{135}_{225}$ (Sect.~\ref{Sect4796A}).

\begin{figure*}
  \includegraphics[trim=2cm 13cm 2cm 2cm, width=15cm]{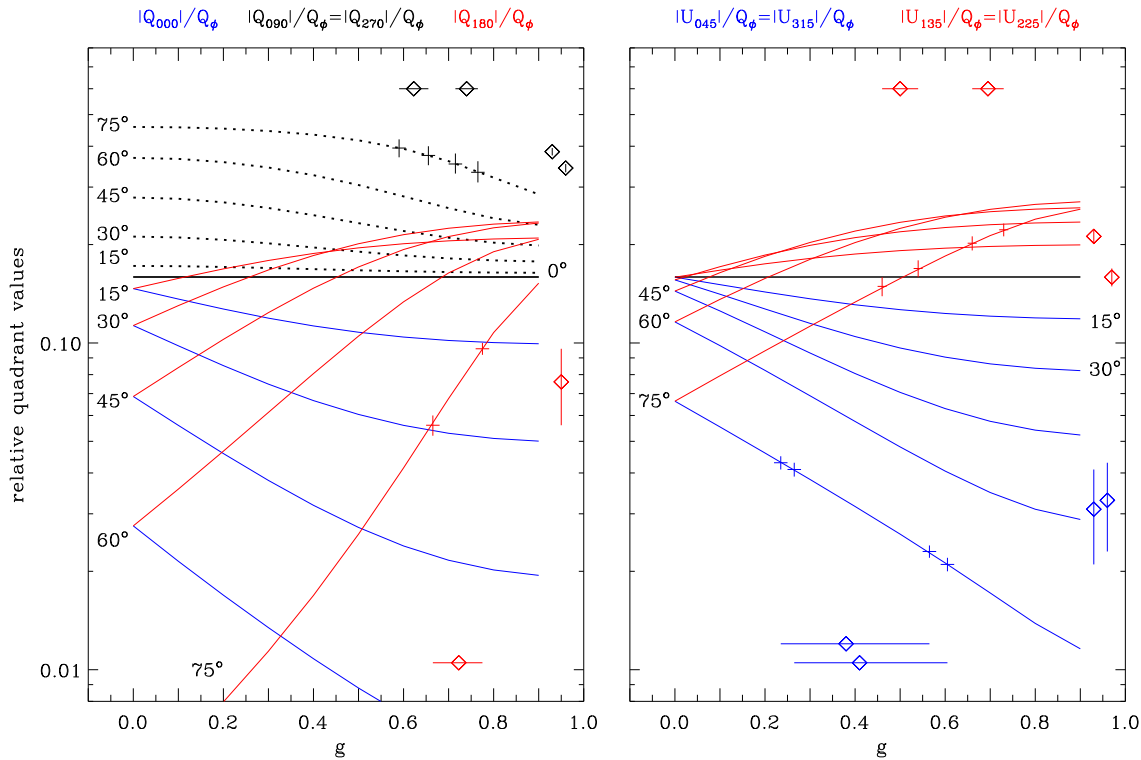}
  \caption{Relative quadrant polarization values $|Q_{000}|/\overline{Q}_\phi$ (blue),
    $|Q_{090}|/\overline{Q}_\phi$ (black) 
    $|Q_{180}|/\overline{Q}_\phi$ (red) on the left and $|U_{045}|/\overline{Q}_\phi$ (blue),
    $|U_{135}|/\overline{Q}_\phi$ (red) on the right for flat disk models with different
    $i$ and as a function of the scattering asymmetry parameter $g$. 
    The measured values for HR 4796A are given on the right side
    in each panel and
    the corresponding $g$-parameter is given at the top or bottom using the
    $i=75^\circ$curves for these derivations.}
\label{FigQuadNorm}
\end{figure*}

\subsection{Relative quadrant parameters}

The azimuthal dependence of the scattering polarization
can be described by the relative quadrant parameters
${Q_{xxx}}/{\overline{Q}_\phi}$ and ${U_{xxx}}/{\overline{Q}_\phi}$.
These parameters are directly
linked to the normalized scattering phase functions
$f_Q^n$, $f_U^n$, and $f_\phi^n$ for the polarization according to
\begin{equation}
  \frac{Q_{xxx}}{\overline{Q}_\phi}=
  \frac{\overline{\epsilon} p_{\rm max} Q_{xxx}^n}
       {\overline{\epsilon} p_{\rm max}\langle f^n_\phi \rangle}
    =\frac{\frac{1}{2\pi}
      \int_{\varphi 1}^{\varphi 2} f_Q^n(\varphi_d,i,g)\,{\rm d}\varphi}
      {\frac{1}{2\pi}
        \int_0^{2\pi} f_\phi^n(\varphi_d,i,g)\,{\rm d}\varphi}
\label{Eqrelquad}      
,\end{equation}
and equivalent to $U_{xxx}/\overline{Q}_\phi$.

The relative quadrant polarization parameters 
$|Q_{xxx}|/\overline{Q}_\phi$ and $|U_{xxx}|/\overline{Q}_\phi$ are plotted
in Figure~\ref{FigQuadNorm} as a function of $g$ for disk inclinations
$i=(0^\circ),\,15^\circ,\,30^\circ,\,45^\circ,\,60^\circ,$ and $75^\circ$.
For pole-on disks, all relative quadrant values are equal to
$(2\,\pi)^{-1}$ independent of the $g$-parameter, because of the
normalization with $\overline{Q}_\phi$. 

With increasing $g$ and $i,$ the quadrant parameters show the
expected steady increase for the front-side quadrants
$|Q_{180}|/\overline{Q}_\phi$ and $|U_{135}|/\overline{Q}_\phi$ (red lines)
and the steady decrease for the backside quadrants
$|Q_{000}|/\overline{Q}_\phi$ and $|U_{045}|/\overline{Q}_\phi$ 
(blue lines). The separation between red and blue lines
produces particularly large
ratios for high inclination because the range of scattering angles
extends from strong forward scattering to strong backward scattering.

For high inclination disks, the substantial contribution of
forward and backward scattering strongly reduces the fractional
scattering polarization in the front-side and back-side quadrants
and the relative quadrant values are therefore
$< (2\,\pi)^{-1}$ for small $g$.  
For the $Q_{000}$ and $Q_{180}$ quadrants, the reduction is further
accentuated by the disk projection which reduces the sampled
disk area for inclined disks. 
Strong forward scattering $g\rightarrow 1$ compensates
these two effects to a certain degree
for the front-side quadrants (red curves), 
while it further diminishes the flux in the (blue)
back side quadrants.

The two positive $Q_{\rm d}$ quadrant parameters $Q_{090}/\overline{Q}_\phi$
and $Q_{270}/\overline{Q}_\phi$ depend mainly on the disk inclination. For
higher inclination, a greater area of the disk is included in these
two quadrants and therefore their relative contribution to the
total polarized flux $Q_{090}/Q_{\phi}$ 
increases from $(2\,\pi)^{-1}$ for $i=0^\circ$ to 0.5 for $i=90^\circ$.
For edge-on systems, there is $2\cdot Q_{090} = \overline{Q}_\phi =
\overline{Q}_{\rm d}$ or all polarized flux of a disk is located only in
the left and right quadrants $Q_{090}$ and $Q_{270}$.  

Figure~\ref{FigQuadNorm} includes the relative quadrant
parameters measured for HR~4796A from Table~\ref{QuadHR4796A}.
This disk has an inclination of about $76^\circ$ \citep[e.g.,][]{Chen20,Milli19}
and the $i=75^\circ$-lines match this value well. For Stokes $Q_{\rm d}$ the
front side value $|Q_{180}|/\overline{Q}_\phi$ is very sensitive for the
determination of the $g$-parameter because the corresponding
curve in Fig.~\ref{FigQuadNorm} is
steep. This results in a value of $g=0.72\pm 0.03$ with a relative
uncertainty of only about $\pm 4~\%$, despite the relatively large
measuring error of about $\pm 13~\%$ for $|Q_{180}|/\overline{Q}_\phi$.
The polarization signal is strong near the major axis and the
corresponding quadrant values $|Q_{090}|/\overline{Q}_\phi$ and $|Q_{270}|/\overline{Q}_\phi$
can be measured with high precision of roughly $\pm 3~\%$.
However, because the $|Q_{090}|/\overline{Q}_\phi$ curve is rather flat
in Fig.~\ref{FigQuadNorm}, the resulting
uncertainty on the derived $g$-values is also about $\pm 3~\%$.
The obtained $g$-values for $|Q_{090}|/\overline{Q}_\phi=0.62$ and
$|Q_{270}|/\overline{Q}_\phi=0.72$ differ significantly because of the described
deviation of the HR 4796A disk geometry from axisymmetry.

For the backside quadrant, only an upper limit of about
$|Q_{000}|/\overline{Q}_\phi<0.04$ could be measured. This limit is not
useful for constraining the $g$-value because expected values
 for an inclination of $75^\circ$ are very low, namely
$|Q_{000}|/\overline{Q}_\phi\approx 0.004$ for $g=0.0$ and $\approx 0.001$ for $g=0.6,$ and the
corresponding curve is below the plot range covered in 
Figure~\ref{FigQuadNorm}.

The relative quadrant values of HR 4796A for Stokes $U_{\rm d}$ 
yield low asymmetry parameters $g\approx 0.4$ for the back-side
quadrants $|U_{045}|/\overline{Q}_\phi$ and $|U_{315}|/\overline{Q}_\phi$, and larger
values of $g\approx 0.5$ and $g\approx 0.7$ for the front side, with
again a significant left--right asymmetry. 

\begin{figure*}
  \includegraphics[trim=2cm 12.5cm 2cm 2cm, width=15cm]{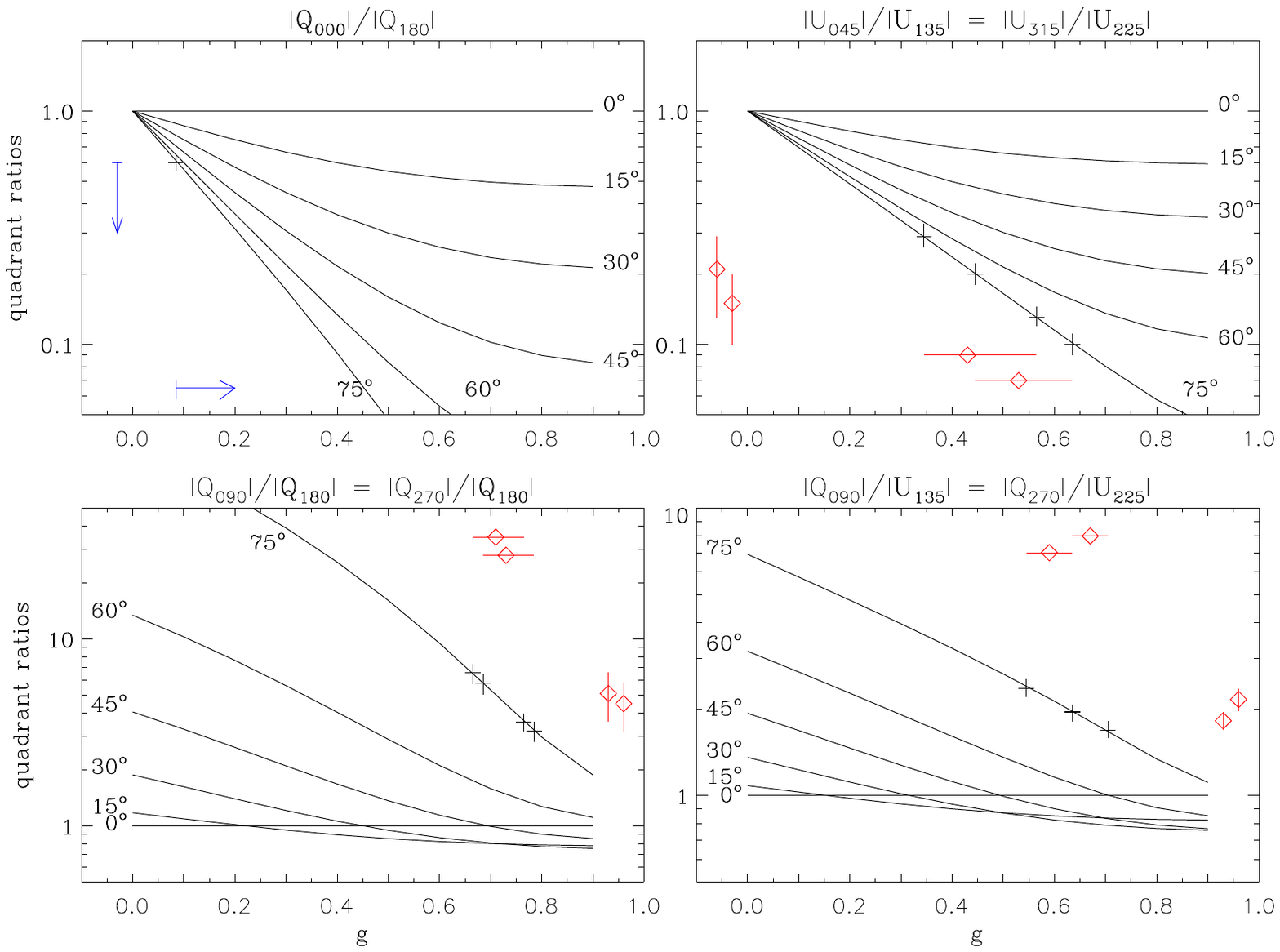}
  \caption{Quadrant polarization ratios for flat disk models for
    $|Q_{000}|/|Q_{180}|$,
    $|U_{045}|/|U_{135}|$, $|Q_{090}|/|Q_{180}|$, and $|Q_{090}|/|Q_{135}|$ with
    measured values from HR 4796A (similar to Fig.~\ref{FigQuadNorm}).
    }
\label{FigRatioQuadrants}
\end{figure*}

\subsection{Quadrant ratios}

The scattering asymmetry $g$ can also be derived from
quadrant ratios describing the brightness contrast between
the front and back sides of the disk as shown 
in Fig.~\ref{FigRatioQuadrants}, which includes 
the measurements from HR~4796A.

High-quality determinations
of $g$ are achieved if the front- and the back-side quadrant polarizations
can be accurately measured. For low-inclination systems,
this should be possible for ratios like $|Q_{000}|/|Q_{180}|$ or
$|U_{045}|/|U_{135}|$ where both the front side and back side
are bright. For high-inclination systems, as in HR 4796A, the back side
can be faint and the
ratios $|Q_{000}|/|Q_{180}|$ or $|U_{045}|/|U_{135}|$ are small
($\lapprox 0.1$) and therefore difficult to measure
accurately. As an alternative, one can use ratios based on the bright
quadrants like $|Q_{090}|/|Q_{180}|$ or $|Q_{090}|/|U_{135}|$ or
equivalent ratios using the right-side quadrants $Q_{270}$ and
$U_{225}$. Many aspects of the diagnostic
diagrams plotted in Fig.~\ref{FigRatioQuadrants} are similar
to the description of the relative quadrant parameters in the
previous section. 

\begin{figure*}
\includegraphics[trim=2cm 13cm 2cm 2cm, width=15cm]{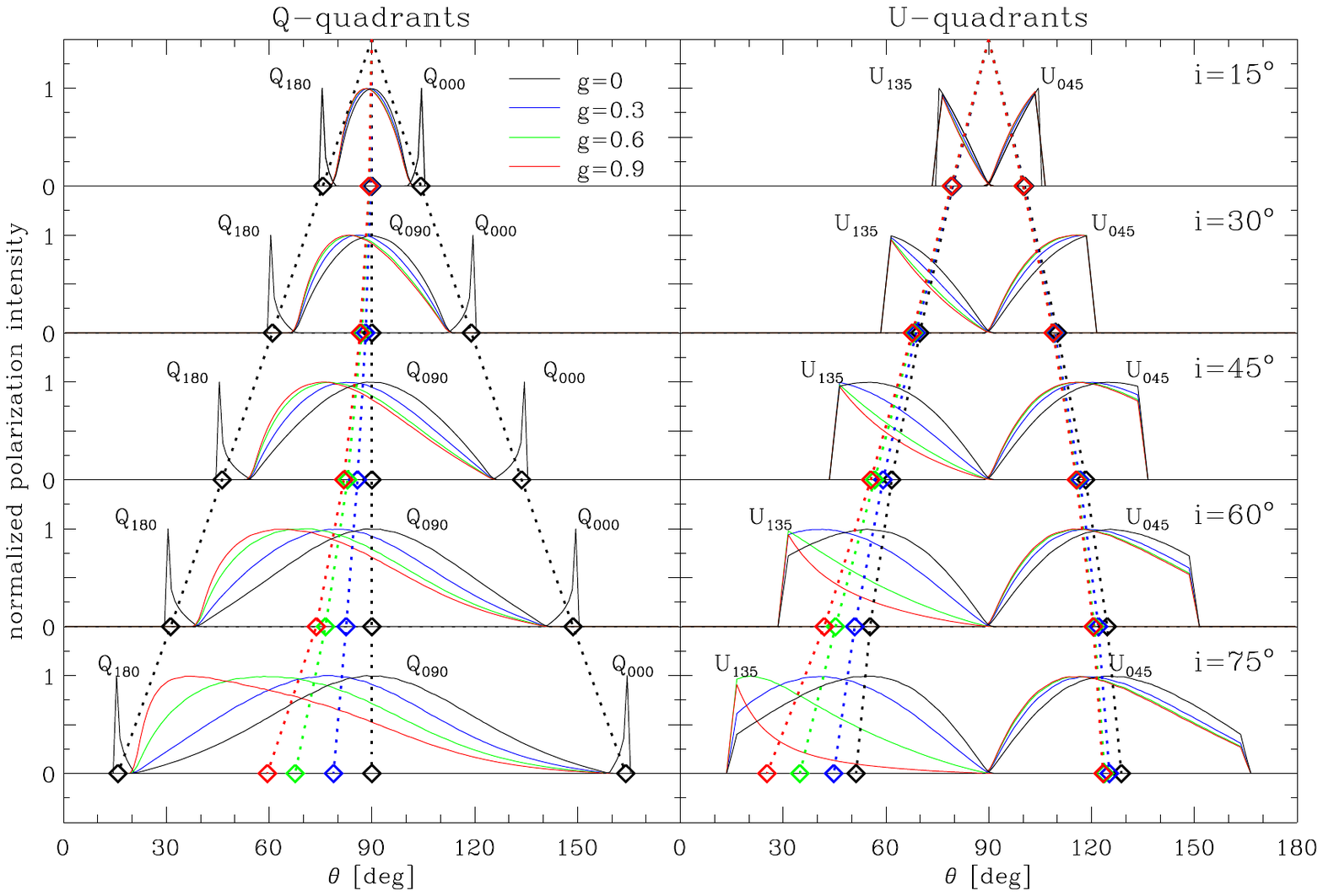}
\caption{Angular distribution of the polarized intensity in the
    polarization quadrants for $i=15^\circ,\,30^\circ,\,45^\circ,\,60^\circ$
    and $75^\circ$ and different asymmetry parameter $g$ (colors).
    For quadrants $Q_{000}$ and $Q_{180}$ only the case $g=0$ is shown,
    because differences for other $g$-parameters are very small.
    The distributions are normalized for each quadrant individually.
    For the quadrants $Q_{000},\,U_{045}$, and $U_{135}$, the median angles
    $\theta_{\rm med}$ (diamonds) for large $g$-values increase
    with inclination as illustrated with the dotted lines.}
\label{ThetaDist}
\end{figure*}

\subsection{Polarized scattering phase function for HR~4796A}

\subsubsection{Comparison of different $g$ determinations}
The measured quadrant polarization parameters for HR~4796A
can be used to strongly  constrain the asymmetry parameter $g$
of the adopted HG$_{\rm pol}$ scattering phase function
$f_\phi(\theta)$, but only for the $\theta$-range sampled
by the used quadrant parameters \citep[see also][]{Hughes18}. The flux-weighted
distribution of scattering angles $\theta$ sampled by a
quadrant strongly depends on inclination $i$, but also on $g$ as
illustrated in Fig.~\ref{ThetaDist}. Indeed, all quadrants
of a nearly pole-on disks probe $f_\phi(\theta)$ only near the
scattering angle of $90^\circ$, while some quadrants probe a
large $\theta$-range for strongly inclined disks. 

We define the median angle $\theta_{\rm med}$ for the angle that
represents the 50th percentile of a cumulative polarized intensity
distribution covered by one quadrant.
The back and front quadrants $Q_{000}$ and $Q_{180}$
sample a narrow range, the median angle
$\theta_{\rm med}$ is close to the most extreme backward
and forward scattering angle $\theta_{\rm med}\approx 90^\circ\pm i$
for a disk with inclination $i$, and $\theta_{\rm med}$ are
essentially identical for different $g$.
The ranges of scattering angles $\theta$ covered by the quadrants
$Q_{090}$, $U_{045}$, and $U_{135}$ are very broad for larger
$i$ and the $\theta_{\rm med}$ depend significantly on
$g$ as shown in Fig.~\ref{ThetaDist}.
For example, $\theta_{\rm med}(Q_{000})$ is $90^\circ$
for isotropic scattering, and becomes smaller for $i\rightarrow 90^\circ$
and $g\rightarrow 1$ as indicated by the colored $\theta_{\rm med}$
points (diamonds) and the dotted lines.

\begin{figure}
\includegraphics[trim=3.0cm 17.5cm 8cm 3.5cm, width=8.8cm]{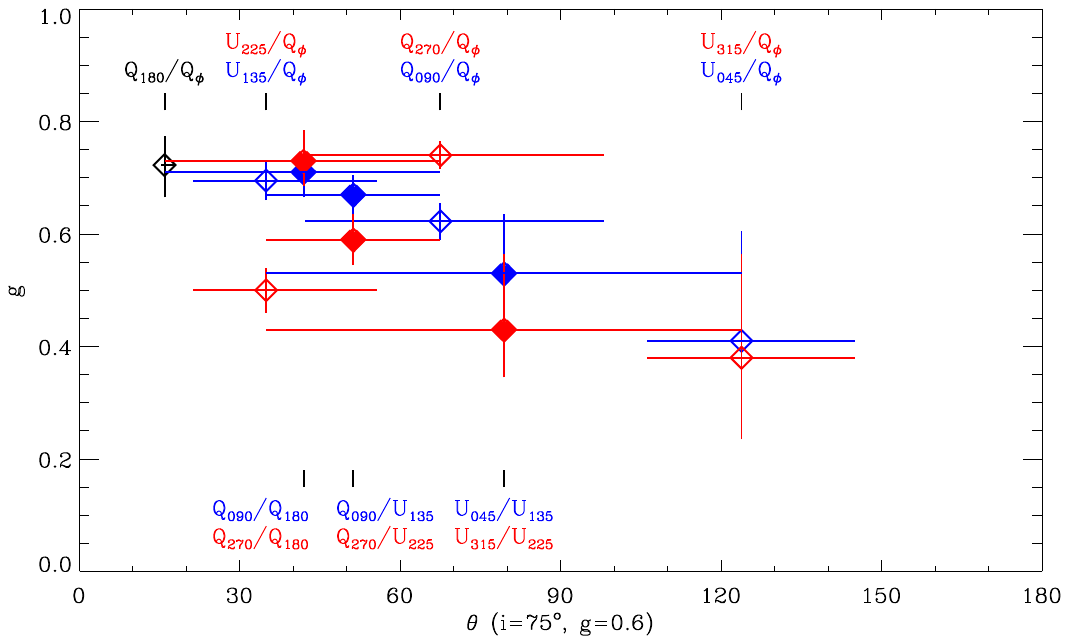}
\caption{Scattering asymmetry parameters for HR~4796A derived from
  relative quadrant parameters (open symbols) and quadrant
  ratios (filled symbols) as a function of the $\theta_{\rm med}$ or
  the mean of $\theta_{\rm med}$, respectively. Colors indicate measurements
  from the left or SW (blue) and the right or NE (red) disk sides.}
\label{FigGTheta}
\end{figure}

The quadrants $U_{045}$ and $U_{135}$ sample the back- and front-side parts of the disk and their $\theta_{\rm med}$-angles
lie between those of the $Q_{\rm d}$-quadrants. The front-side quadrant 
also shows a strong tendency towards smaller $\theta_{\rm med}(U_{135})$ values
for larger $g$ and $i,$ as in the $Q_{090}$-quadrant, while the
$g$-dependence of $\theta_{\rm med}(U_{045})$ is much smaller.

Figure~\ref{FigGTheta} shows the $g$-parameters obtained for HR~4796A as
a function of the $\theta$-angle probed by the used quadrant parameters.
The angle $\theta$ corresponds for relative quadrant parameters to
$\theta_{\rm med}$ for $i=75^\circ$ and $g=0.6$ as
given in Fig.~\ref{ThetaDist} and the horizontal uncertainty bar
spans two-thirds of the plotted $\theta$-distribution
(from the 16.6 to the 83.3 percentiles). For the
quadrant ratios, the adopted $\theta$-values are the mean 
of the $\theta_{\rm med}$ of the two quadrants and the horizontal bars illustrate
their separation.  In principle, one should
consider for the $\theta_{\rm med}$-values the systematic trend
of $g(\theta_{\rm med})$ from higher $g$-values ($\approx 0.7$)
for forward-scattering
quadrants to lower values ($\approx 0.4$) for the backward
scattering quadrants. We neglect
this effect which would introduce $\theta_{\rm med}$ shifts of about
$-5^\circ$ for $U_{135}$ and $U_{225}$, shifts of about $+5^\circ$ for $Q_{090}$ and
$Q_{270}$, and smaller shifts for the other quadrants. 

Figure~\ref{FigGTheta} shows for HR 4796A a systematic dependence
of the derived $g$-parameters with scattering angle $\theta$.
The results from the relative quadrant parameters
and the quadrant ratios are roughly consistent. 
The colors indicate measurements for the left (blue) and right (red)
sides of the disk and $g$-values differ significantly
between the two sides
for $\theta_{\rm med}=35^\circ$ and $70^\circ$ because of the
left--right disk brightness asymmetry.  On the fainter side, this effect 
reduces the derived $g$-value for $U_{225}/\overline{Q}_\phi$,
while $g$ is enhanced for $Q_{270}/\overline{Q}_\phi$ because the
intrinsic faintness of $Q_{270}$ mimics a disk with relatively little
$90^\circ$-scattering because of the normalization with $\overline{Q}_\phi$.
The intrinsic left--right brightness asymmetry of HR 4796A has
less impact on the $g$ determination based on quadrants ratios from the
same side. This redundancy helps to disentangle
the effects of the scattering asymmetry $g$ from 
geometric or left--right disk brightness asymmetries.

\begin{figure}
\includegraphics[trim=2.5cm 13.5cm 7cm 2.5cm, width=8.8cm]{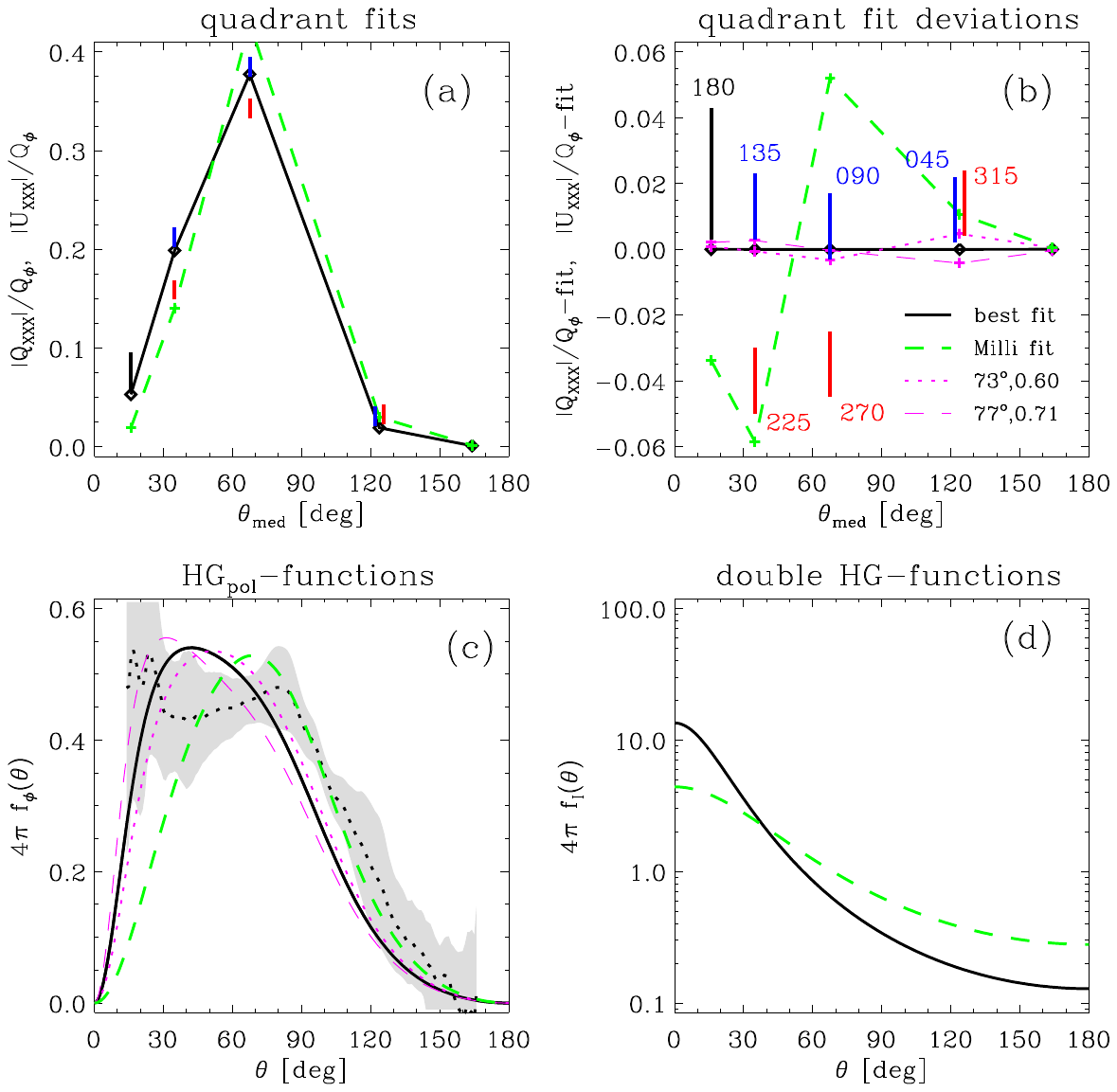}
\caption{(a) Comparison of the best single parameter
  HG$_{\rm pol}$ fit model for fixed $i=75^\circ$ and measured relative quadrant
  polarization parameters for HR 4796A. (b) Deviations from the best-fitting HG$_{\rm pol}(i=75^\circ,g=0.65)$ for the
  measured values, for the best fits for slightly different disk inclinations
  HG$_{\rm pol}(i=73^\circ,g=0.60)$ and HG$_{\rm pol}(i=77^\circ,g=0.71)$
  and for the HG$_{\rm pol}$ fit from \citet{Milli19} (green dashed line).
(c) HG$_{\rm pol}$ phase functions and the directly extracted
  phase function $f_\phi(\theta)$ (dotted black line) with corresponding
  uncertainty range (grey shaded area) from \citet{Milli19}. (d) HG intensity phase functions. The measured quadrant values and
  the different fit curves are identified in panel (b).}
\label{QuadFitHG}
\end{figure}

\subsubsection{A ``mean'' asymmetry parameter $g$ for HR~4796A}
\label{Sectmeang}
The clear trend of the derived $g$-parameter with scattering
angle $\theta$ for HR~4796A in Fig.~\ref{FigGTheta} reveals
that the used HG$_{\rm pol}$-function is an oversimplified description
of the polarized scattering phase function for this object.
For fainter or less well resolved disks, and for those with low inclination, it may not be possible to recognize such systematic
deviations from a HG$_{\rm pol}$ function, and for all these
cases the derived $g$-value from the HG$_{\rm pol}$-function
could serve as a good starting point
for the analysis of quadrant polarization parameters. 

Therefore,  for HR~4796A we  also derive a ``mean'' value for the
HG asymmetry parameter $g$ despite the discussed trend.
To this end, for the seven measured relative quadrant
values $|U_{045}|/Q_\phi$, $|Q_{090}|/Q_\phi$, $|U_{135}|/Q_\phi$,
$|U_{180}|/Q_\phi$, $|U_{225}|/Q_\phi$, $|U_{270}|/Q_\phi$, and $|U_{315}|/Q_\phi$
from Table~\ref{QuadHR4796A},  we determine the best-fitting $g$-asymmetry parameter
for the adopted disk inclination $i=75^\circ$.
This yields HG$_{\rm pol}(i=75^\circ,g=0.65)$ with a weighted sum of squared
deviations of $\chi^2=15.4$, and the corresponding calculated and
measured values are plotted
in Fig.~\ref{QuadFitHG}a. The differences between the models 
are more visible in Fig.~\ref{QuadFitHG}b, where
the deviations of data points and calculations from the best-fit
model are shown. The large $\chi^2$-value indicates
that the adopted HG$_{\rm pol}$-fit does not describe  the data well because
of the significant left--right asymmetry between $|Q_{090}|$ and $|Q_{270}|$,
or $|U_{135}|$ and $|U_{225}|$ at the level of about $4~\sigma$ ($\sigma$:
standard deviations), which cannot be described with an axisymmetric
disk model. Additionally, the best fit underestimates the quadrant values at
small ($\theta_{\rm med}=16^\circ$) and large ($124^\circ$) scattering angles.
Panel (b) also includes the best-fit results for slightly different
disk inclinations HG$_{\rm pol}(i=73^\circ,g=0.60)$ and
HG$_{\rm pol}(i=77^\circ,g=0.71)$, which differ very little and produce
deviations between fit and data that are  similar to the $i=75^\circ$ solution.
The corresponding Henyey-Greenstein scattering phase functions
HG$_{\rm pol}$ (or $f_\phi(\theta)$) are given
in panel (c), while panel (d) shows the HG intensity function ($f_I(\theta)$) for
the best quadrant solution for $i=75^\circ$ and
the solution from \citet{Milli19}. 

Fortunately, we can compare the result from the
quadrant parameter fitting with the analysis of the same HR~4796A data by
\citet{Milli19}. They extracted from the polarimetric imaging data
a detailed phase curve shown in Fig.~\ref{QuadFitHG}c
covering the scattering angle range
$\theta=13^\circ$ to $145^\circ$. They  also fitted the extracted
phase curve with a single parameter HG$_{\rm pol}$-function and
obtained an asymmetry parameter of $g=0.43$ which is much smaller than our
value of $g=0.65$ (Fig.~\ref{QuadFitHG}c). An important reason
for this discrepancy is the sampling of the scattering angles of the
data used for the fitting. In this work, the
fitting is based on seven quadrant polarization values ---for a disk with $i=75^\circ$ and $g>0.4$--- which are strongly biased towards
small $\theta$-values because of the forward ``distorted''
distribution of the polarized flux.
In addition, the back-side
quadrants $U_{045}$ and $U_{315}$ are weak and the corresponding
measurements have a low signal-to-noise ratio of
$S/N< 5$ and therefore a small weight. Thus, the fit to the
quadrant values predominantly samples the range
$\theta\approx 16^\circ$ to $67^\circ$ of the
scattering phase function. The analysis of \citet{Milli19} samples
a much broader range and particularly also more
backward scattering angles. Therefore, the result of these latter authors of $g=0.43$
closely matches the $g$-values derived in this work by the quadrant ratios
$|U_{045}|/|U_{135}|$ and $|U_{315}|/|U_{225}|$ (Fig.~\ref{FigRatioQuadrants}).
On the other hand, the HG$_{\rm pol}$-fit of \citet{Milli19}
 underestimates their extracted phase function in the
forward-scattering range $\theta\approx 16^\circ$ to $35^\circ$.
This comparison illustrates the bias effect that can be
introduced by different kinds of phase
curve sampling, if the adopted model curves $f_\phi(\theta)$
do not match well the real scattering phase function of the dust.

\begin{figure}
  \includegraphics[trim=2.5cm 13.5cm 7cm 2.5cm, width=8.8cm]{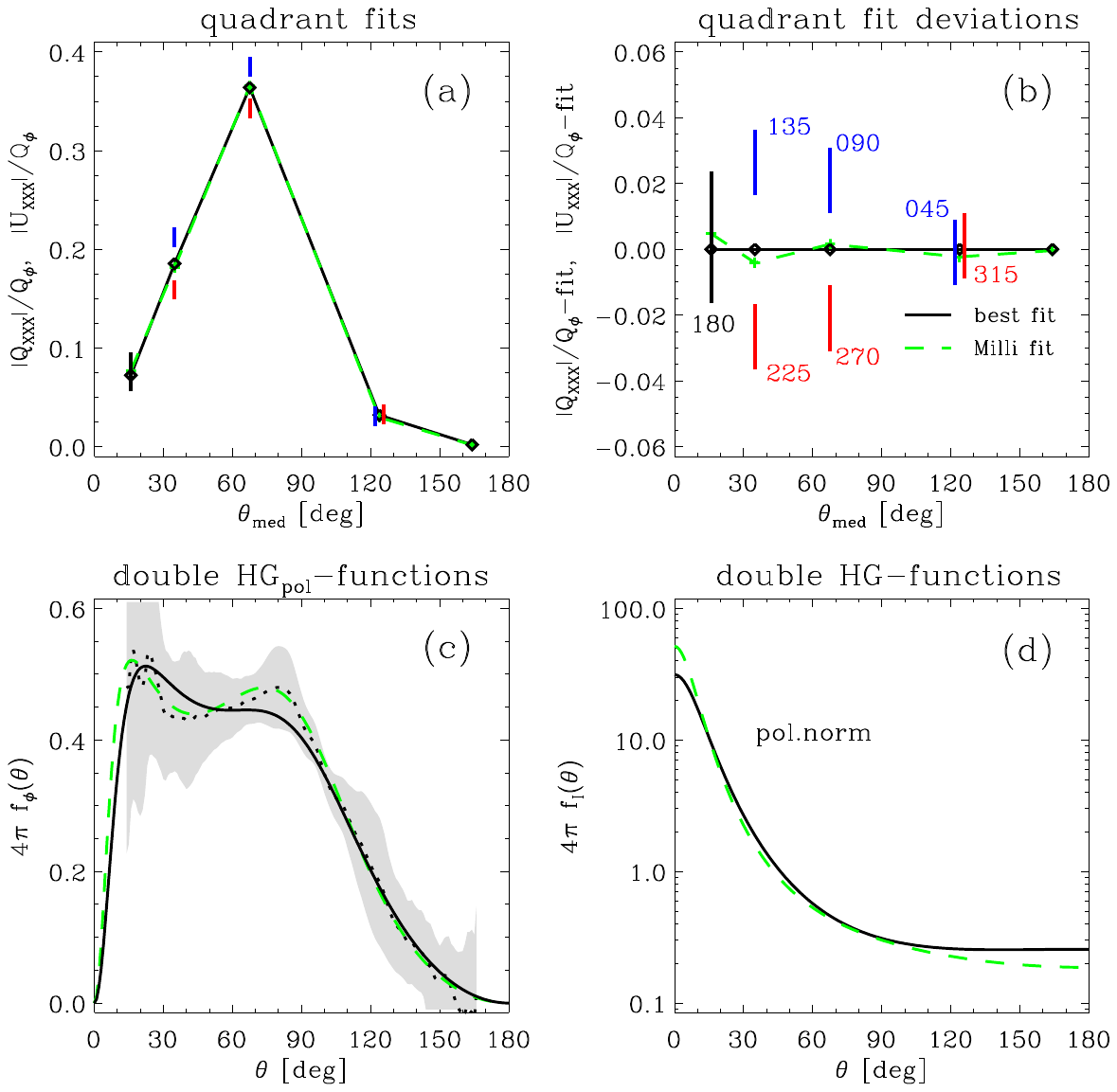}
  \caption{Same as Fig.~\ref{QuadFitHG} but for the best-fitting
    double HG$_{\rm pol}$ scattering phase function
    $(g1,g2,w)_A=(0.78,-0.09,0.85)$ for $i=75^\circ$ and the function
    obtained by \citet{Milli19} (dashed green lines):
(a) calculated and measured relative quadrant values;
    (b) deviations of obtained values from the best-fitting solution;
    (c) double HG$_{\rm pol}$ phase function for the
    fits and the directly extracted curve from \citet{Milli19} (dotted
    line and gray uncertainty range);
(d) corresponding intensity phase functions.}
\label{QuadFitDoubleHG}
\end{figure}

\subsubsection{A fit with a double HG$_{\rm pol}$-function}

A better agreement between measured and calculated
quadrant values can be obtained using a double HG-function for the
dust scattering
\begin{equation}
f_I(\theta,g_1,g_2,w)=w f(\theta,g_1) + (1-w) f(\theta,g_2),
\end{equation}
because three parameters provide more freedom for the
polarized phase curve
$f_\phi^{\rm n}(g_1,g_2,w)=f_I(\theta,g_1,g_2,w)\cdot p_{\rm sca}(\theta)$
in the quadrant fitting.
Calculating quadrant polarization values for $i=75^\circ$ and
a grid of phase function parameter  
$g_1$ and $w \in [0.00,0.01,...,1.00]$, and
$g_2\in [-1.00,-0.99,...,g_1]$ gives a best-fit solution of
$(g1,g2,w)=(0.78,-0.09,0.85)$ with $\chi^2=10.2$ which is plotted in 
Fig.~\ref{QuadFitDoubleHG}. 

Compared to the case of the best single parameter HG$_{\rm pol}$ function
(Fig.~\ref{QuadFitHG}), this fit does not underestimate the relative
quadrant values at $\theta_{\rm med}=16^\circ$ and $124^\circ$
and passes in the middle of the discrepant quadrant
values at $35^\circ$ and $67^\circ$ for the left and right disk sides
(see Fig.~\ref{QuadFitDoubleHG}a and b). The corresponding
polarized scattering phase function $f_\phi(\theta)$ in
Fig.~\ref{QuadFitDoubleHG}c
has a much wider peak extending from $\theta\approx 20^\circ$
to $\approx 90^\circ$,  closely matching the directly extracted
phase curve from \citet{Milli19}.

It is interesting to compare our results with the double HG$_{\rm pol}$ function
obtained by \citet{Milli19} from the fit to the detailed phase-curve extraction which is included in Fig.~\ref{QuadFitDoubleHG}
as a green dashed line. Unfortunately, there is an error in the
indicated fit parameters in Fig.~5
of \citet{Milli19} but the plotted fit curve is correct. The
fit parameters should be
$(g1,g2,w)_{\rm Milli}=(0.83,0.09,0.81)$ (J. Milli, personal communication)
which also provide a very good fit to the quadrant
values derived in this work as shown in Fig.~\ref{QuadFitDoubleHG}(a) and (b).

The good agreement between the double HG$_{\rm pol}$-fits
  of \citet{Milli19} and the solution found for the quadrant polarization
  values
shows that the selection of a more appropriate scattering phase function strongly
reduces the large difference in the deduced $g$-determination
described in Sect.~\ref{Sectmeang} using only the single HG$_{\rm pol}$ function.    
It should also be noted that the polarized phase function
fit $f_\phi(\theta)$ of \citet{Milli19} does not consider the left--right
asymmetry of the disk in HR~4796A and therefore the phase
curve uncertainty attained by these latter authors is larger than their measurement uncertainties.
It seems likely that the azimuthal polarization signal extracted
by \citet{Milli19} would probably allow the determination of a
better constrained empirical $f_\phi(\theta)$-function for HR~4796A if
the significant azimuthal dependence on the dust density 
is included in the fitting. Using a more detailed disk model for
the fitting of the derived quadrant polarization parameters seems
to be less useful because of the small number of measured values, a matter that is discussed further in Sect.~\ref{Sectlimits}.

This example shows that selecting a good model fit function is
important for the analysis of the quadrant polarization parameters
and this should be investigated
in more detail. The double HG$_{\rm pol}$ is probably not an ideal
choice, because in the intensity scattering function 
significant weight is given to the forward and backward scattering angles,
which produce less polarization and therefore contribute less to the signal in the corresponding polarization quadrants
$Q_{000}$ or $Q_{180}$. This could explain the substantial differences
for $\theta<16^\circ$ or for $\theta>124^\circ$ between
the two derived best-fitting functions in Fig.~\ref{QuadFitDoubleHG}d
where the quadrant parameters provide no or only weak constraints on
the shape of the scattering phase function.

In a future study, alternative
polarized scattering phase functions $f_\phi(\theta)$
should be investigated for the fitting of polarimetric data, which give more weight
to intermediate scattering angles $\theta\approx 90^\circ \pm i$. 
Such a curve should also consider deviations of the fractional
scattering polarization from a symmetric curve (Rayleigh-like)
with respect to $\theta=90^\circ$ as already derived from
observations of HR~4796A by \citet{Perrin15} and \citet{Arriaga20}.
Considering this could be particularly important when
constraining $f_\phi(\theta)$ for the dust in debris disks with
smaller inclinations and a more limited
observable range of $\theta$-angles.

\section{Discussion}

\subsection{New polarization parameters for circumstellar disks}

In recent years, the scattering light of many proto-planetary and debris disks has been spatially resolved with high-resolution
polarimetric imaging using modern AO systems at large telescopes
\citep{Schmid21}.
Unfortunately, the presented results for the measurements of the
polarized light from circumstellar disks
are highly heterogeneous and are rarely flux calibrated, and
it is therefore very difficult to compare the results from different studies
for a systematic investigation of disks.

The main motivation of the present paper is the promotion of a photo-polarimetric
parameter system which should help to homogenize the polarimetric
measurements for circumstellar disks
and allow more
straightforward comparisons between measurements of different disks
and model results. The introduced quadrant polarization
parameters $Q_{000}$, $Q_{090}$, $Q_{180}$, $Q_{270}$ and $U_{045}$,
$U_{135}$, $U_{225}$, $U_{315}$ are defined for the Stokes $Q_{\rm d}$ and $U_{\rm d}$
parameters aligned with the apparent major and minor axis of the
projected disk; they are based on the ``natural'' quadrant pattern
produced by circumstellar scattering and measure within these
quadrants the integrated Stokes $Q_{\rm d}$ and Stokes $U_{\rm d}$ flux, respectively.

These eight quadrants are very well suited for the description
of the azimuthal dependence of the polarization signal of disks,
except for edge-on or nearly edge-on systems.
Furthermore, they can be used to quantify geometric deviations of the disk from axisymmetry
from differences between left and right quadrants  
or characterize the disk inclination effects and the dust scattering
asymmetry from ratios between back-side and
front-side quadrants.

This disk characterization only requires differential
polarization measurements, like relative quadrant parameters
$Q_{xxx}/\overline{Q}_\phi$ and $U_{xxx}/\overline{Q}_\phi$, or quadrant
ratios like $Q_{000}/Q_{180}$. No absolute flux calibration with
respect to the intensity of the star $I_{\rm star}$ or the disk
$\overline{I}$ are required and therefore
one can also use polarimetric imaging of disks obtained in coronagraphic mode
or with the central star saturated. In addition, the eight quadrant
parameters are partially redundant and offer multiple options for the
characterization of a disk, meaning that
problems with a particular quadrant, for example because of the peculiarities
of a disk or observational effects, can be mitigated.

The quadrant polarization measurements should be particularly
well adapted for well-resolved, extended, low-surface-brightness
debris disks, which are relatively common \citep{Esposito20}.
The integration of the $Q$ and $U$ polarization for entire quadrants helps
to improve the signal, and restricting the measurements of
the azimuthal dependence to a few values is appropriate
for a faint source where it is hard to get sufficient signal for
a detailed characterization. Of course, the calibration of the
polarization zero point must be determined very accurately for
faint sources and this can be achieved for many debris disks
because the central star is often a very good zero polarization
reference source.

Model calculations exploring the parameter space for
the dust scattering in circumstellar disks are of particular importance for
advancing our understanding of the properties of the scattering
dust in disks. The quadrant polarization parameters are 
very well suited to characterizing the azimuthal dependence of the
polarization signal for different models. Because these
model results can be expressed as relative values or ratios, they
can be readily compared with each other for the evaluation
of dependencies on the scattering asymmetry for optically thin
disks or the angle dependencies of the surface reflectivity in optically
thick disks, even if parameters such as stellar illumination, disk size, or radial dust density distribution in
optically thin disks are different.

The modeled values can also be compared with
observations, but important issues
are the PSF smearing and
polarimetric cancelation effects between positive and negative
quadrants. This can significantly reduce the measurable polarization
for poorly resolved disks \citep{Schmid06,Tschudi21} and change
the appearance of the $Q$ and $U$ quadrant patterns of inclined
or asymmetric disks \citep{Heikamp19}.
For example, for the Stokes $Q_{\rm d}$ quadrants, the PSF convolution
reduces the total signal of the positive quadrants $Q_{090}$ and $Q_{270}$
by the same amount as it enhances the signal (less negative signal) in the
negative quadrants $Q_{000}$ and $Q_{180}$ ; the same is true  for the Stokes $U_{\rm d}$
quadrants.
If the PSF is well known for a given
observation then the smearing and cancelation effects can
be taken into account accurately in order to minimize
the introduced effects \citep{Tschudi21}.

The measurements of the quadrant polarization parameters
for circumstellar disks
provide a simple and model-independent method for the
description of the azimuthal distribution of the scattering
polarization and the obtained results can be easily
tested by comparing the measured and 
calculated model values. Similarly, the quality of the measured quadrant
data can be verified with alternative measurements of the same target.
Of course, a detailed analysis
of the polarimetric imaging data with 2D synthetic model images
would provide a more detailed comparison,
but this is a very laborious procedure which
requires detailed knowledge of the observational effects for
each data set and a good understanding of the modeling aspects
for each individual disk
\citep[see e.g.,][for the case of HR 4796A]{Milli19,Olofsson20,Chen20}. 
Therefore, it appears attractive to base a quick analysis of many
disks on the simple quadrant polarization parameters. Once measured
and corrected for the PSF smearing, they remain unchanged
until higher quality measurements become available and the
interpretation of the measurements obtained can be continuously
improved if additional information about the corresponding disk
model can be taken into account.

\subsection{Investigation of debris disks}
The usefulness of the quadrant polarization parameters
is tested in this work with simple models of debris disks and
with observations of the prototype debris-disk system HR~4796A.
Debris disks are optically thin and therefore the azimuthal
dependence of the polarization signal depends directly
on the polarized scattering phase function $f_\phi(\theta)$
of the dust. Because the quadrant polarization parameters
measure the azimuthal dependence, they are ideal
for determining $f_\phi(\theta)$. This is shown with model
calculation of flat axisymmetric debris disks using
the simple HG$_{\rm pol}$ function for the parameterization of the
polarized scattering phase function $f_\phi(\theta,g)$.
For optically thin, rotationally symmetric disks, the azimuthal
dependence of the polarization signal can be directly described by
a disk scattering phase function $f_\phi(\varphi_d,i,g)$, which only
depends on the disk inclination $i$ and the scattering asymmetry parameter
$g$ of the HG function. This function also defines the relative
quadrant polarization values $Q_{xxx}(i,g)/\overline{Q}_\phi(i)$
as an eight-parameter condensation of the azimuthal polarization
dependence, from which one can also derive quadrant ratios like
$Q_{000}(i,g)/Q_{180}(i,g)$ as alternative results. These 
parameters yield a measure for the dust-scattering
asymmetry  for a given $i$  and corresponding
diagnostic diagrams have been calculated for
relative quadrant values and quadrant ratios.
If the selected HG$_{\rm pol}$ function is an appropriate
parametrization for the dust scattering of an
observed debris disk then all the measured quadrant parameters
should yield the same $g$ parameter. The same method can be
applied for investigations of other scattering phase functions.

We tested the polarized phase-curve determination based on the
quadrant polarization parameters for data of the  ``prototype''
debris disk around HR~4796A from \citet{Milli19}. First, we
noticed a significant disk asymmetry between
the ``left'' and ``right'' sides with respect to the minor axis
of the disk ring as projected on the sky. We did not
consider this disk asymmetry and simply derived a ``mean''
scattering phase curve accepting that this introduces
some uncertainties in the phase curve analysis.
The diagnostic diagrams
for the HG$_{\rm pol}$ scattering phase function were used and
the obtained $g$ parameter determination shows a clear trend
from high values $g\approx 0.7$ for quadrants sampling
small scattering angles $\theta\approx 30^\circ$ to lower values
$g\approx 0.4$ for larger scattering angles $\theta\approx 120^\circ$. 
This is a clear indication that the adopted HG$_{\rm pol}$-scattering
function is not adequately describing the dust in HR~4796A.
The over-simplified fit model introduces strong bias effects
responsible for significant differences between the $g$-value
determination based on a detailed phase-curve extraction and
the one based on the quadrant parameters.

As alternative, we used a three-parameter double HG$_{\rm pol}$ function
as description for the dust scattering $f_\phi(\theta)$.
The best fit solution to the quadrant values
that we find is in good agreement with the
detailed phase curve extraction of \citet{Milli19} based on the
same data. This is surprising because the covered $\theta$-range
for the phase curve from about $16^\circ$ to $164^\circ$ 
is large for the high-inclination ($i\approx 75^\circ$)
system HR~4796A and a characterization of $f_\phi(\theta)$ based on
a few quadrant values yields only a relatively coarse $\theta$ resolution.
The main reason the detailed $f_\phi(\theta)$-extraction
of \citet{Milli19} is not clearly superior when compared 
to the quadrant method is the significant deviations of the
dust density distribution from axisymmetry, which were also
not taken into account by \citet{Milli19} for their phase-curve
fitting. The detailed extraction contains much more
information on the nonsymmetric disk brightness distribution
which is discussed in detail in \citet{Milli19}. Also,
small-scale structures are seen in the extracted azimuthal polarization
curve of HR~4796A for which the quadrant parameters
are ``blind''.

However, the HR~4796A example shows that the analysis on
the quadrant polarization
parameters performs rather well for high-quality data of a bright target
if we are ``only'' interested in the global azimuthal polarization
dependence of debris disks caused by the polarized scattering phase
function $f_\phi(\theta)$ of the dust.

\subsection{Limitations}
\label{Sectlimits}

The quadrant polarization parameters are designed for
a simple description and analysis of the azimuthal dependence of the
scattering polarization of
circumstellar disks. The method is well defined but it has
limitations, which must be taken into account in the
interpretation of the results. 

Importantly, one should be aware that real disks
are often quite complex and a description using only 
eight parameters or less  only yields rough information
about the left--right disk asymmetry and the differences
in the front--back brightness distribution.
There are various effects that can cause significant
departures from axisymmetry in the disk geometry:
an intrinsic ellipticity introduced by noncircular orbits of
dust particles, different types of hydrodynamic instabilities
introducing spiral structures, azimuthal density features,
lobsided disks, shadows cast by unresolved dust structures
near the central star, dynamical interactions with proto-planets or other
gravitating bodies in the system, and probably other effects.

Such asymmetries can be identified easily as left--right differences
but they can also produce brightness
effects between the disk front- and backside which are then
blended with front--back brightness effects caused
by the dust-scattering asymmetry in optically thin disks or the
angle-dependent surface reflectivity in optically thick disks.

For a disk with
complex morphology  based on a small
number of measured polarization quadrant parameters, it can be difficult  to recognize whether
the asymmetries are caused by the disk geometry,
the scattering phase function, optical depth effects,
or an observational problem. Therefore, it
is certainly always useful to examine the disk polarization
images for the presence of strong azimuthal structures which
can be taken into account for disentangling the effects of
the disk geometry from those of the dust scattering phase function
for an interpretation of the data.

Using only eight quadrant polarization parameters for the
characterization of the detailed structure of a well-observed disk can of course only provide limited information as
demonstrated for HR~4796A. In such a case, a detailed extraction of the
polarized flux or a two-dimensional model
fitting to the data as in \citet{Milli19} or \citet{Arriaga20}
will provide more accurate results and a less ambiguous interpretation.
For example, a detailed model analysis for HR~4796A
could consider two or three parameters for 
the intensity scattering phase function $f_I({\theta})$,
one or two parameters for the shape of the fractional scattering
polarization $p_{\rm sca}(\theta)$ (and not only the fixed Rayleigh
scattering like curve given in Eq.~\ref{Eqpsca}), a description
of the ring geometry, and three or more parameters for the
azimuthal dust density distribution.

Therefore, the quadrant polarization parameters are less suitable for
a detailed investigation of well-observed disks, and are more suitable for the
exploration and the approximate description of the global properties of the scattering
dust in many disks. However, it is still useful to derive these
parameters for well-observed disk prototypes for a comparison
with disks for which a detailed analysis is hardly possible,
or for multi-wavelength studies of a given disk where a few well-defined parameters are sufficient to recognize and quantify wavelength
dependencies for the polarized dust scattering phase function.

\subsection{Conclusions}

The quadrant polarization parameters introduced
in this work seem to be very useful for a simple description
of the azimuthal dependence of the polarization signal of
circumstellar disks. These parameters can be
determined from observations of many different types
of circumstellar disks, for example debris disks around
young or old stars, with or without strong illumination or
dust blow-out signatures, or for proto-planetary disks with
small or large central cavities and different kinds of
hydrodynamical features.

The measured quadrant parameters can be compared with disk models
that take the PSF smearing and cancelation effects into account
and can explore the expected polarization signatures
introduced by different descriptions for the scattering dust.
Accumulating such data for a larger sample will allow
a search for systematic trends in dust scattering properties
for different disk types and for different wavelengths and inform
us about the homogeneity or heterogeneity of dust-scattering
properties in circumstellar disks. This can be achieved with relatively
small uncertainties when compared to circumstellar shells or clouds,
because the scattering angles $\theta$, which
have an important impact on the produced polarization signal,
are typically very well known for resolved circumstellar disks. 
Investigations of the dust in circumstellar disks are also
very attractive because many studies indicate that the dust
evolves strongly in these systems and this could produce
systematic trends for different disk types, which could be measurable
with the new generation of AO polarimeters. 

\begin{acknowledgements}
  I am very grateful to Julien Milli for the reduced $Q_\phi$ and
  $U_\phi$ images of HR 4796A used in this study, for the polarized
  scattering phase curves derived in \citet{Milli19}, and for many useful
  comments on an earlier version of this manuscript. I am indebted to
  an anonymous referee who made a very detailed and thoughtful
  review of the submitted manuscript which helped to improve the final
  paper significantly. I also thank Jie Ma for a careful reading of
  the manuscript and for checking the mathematical formulas.
  This work has been carried out
  within the framework of the National Center for Competence in
  Research PlanetS supported by the Swiss National Science Foundation.
\end{acknowledgements}

\bibliographystyle{aa} 
\bibliography{DiskQuadrants5.bib} 


\begin{appendix}
\section{Radiation parameters for debris disks.}
The following IDL procedure calculates disk-averaged scattering
functions and normalized quadrant polarization parameters for
flat, rotationally symmetric, and optical thin disks
with HG and HG$_{\rm pol}$ scattering phase functions.
Input parameters are the disk inclination in degrees
{\verb|ideg|} $\in [0^\circ,90^\circ]$ and the scattering
asymmetry parameter {\verb|g|} $\in\, ]-1,+1[$. Output parameters
are {\verb|fiavg|} for the disk averaged intensity scattering phase function
$\langle f_I(i,g) \rangle$ (Eq.~\ref{Eqintphir}) and {\verb|fphiavg|} for
the corresponding normalized function for the azimuthal polarization
$\langle f_\phi^n(i,g) \rangle$ (Eq.~\ref{Eqintqphi}).
In addition, the procedure provides the five normalized
quadrant polarization parameter {\verb|qpp[0], qpp[1], qpp[2], qpp[3],|} and {\verb|qpp[4]|} corresponding to
$Q^n_{000}(i,g)$, $U^n_{045}(i,g)$, $Q^n_{090}(i,g)$, $U^n_{135}(i,g)$ and
$Q^n_{180}(i,g)$, respectively (Sect.~\ref{NormQuad1}). The Stokes $Q_{\rm d}$ phase
function follows from the quadrant sum
$\langle f_Q^n(i,g) \rangle= 2\cdot Q^n_{180}+Q^n_{000}+Q^n_{090}$.

\begin{table*}
  \caption{IDL procedure for the calculation of disk-averaged scattering
functions and normalized quadrant polarization parameters.}  
\begin{tabular}{ll}
{\verb|pro quadrants,ideg,g,fiavg,fphiavg,qpp|\hfil} \\
{\verb|incl = ideg*!pi/180.| \hfil}        & \hspace{-3cm}; inclination $i$ in radians \\
\noalign{\smallskip}
{; array of disk azimuth angles $\varphi_d$ for disk ring: 3600 point with [0,\,0.1,\,..\,,\,359.9] degrees in radians
  \hfil }\\
\noalign{\smallskip}
{\verb|phi_d = findgen(3600)*!pi/1800.|\hfil}  \\
{\verb|x = -sin(phi_d)| \hfil}            & \hspace{-3cm}; x-sky for inclined disk ring \\
{\verb|y = cos(incl)*cos(phi_d)|}        & \hspace{-3cm}; y-sky   \\
{\verb|z = -sin(incl)*cos(phi_d)|\hfil}  & \hspace{-3cm}; z along line of sight \\
{\verb|phi = atan(-x,y)|}                  & \hspace{-3cm}; sky azimuth angle $\phi_{xy}(\varphi_d,i)$ (Eq.~\ref{Eqphiphi})  \\
{\verb|theta = acos(z)|\hfil}              & \hspace{-3cm}; scattering angle $\theta(\varphi_d,i)$ (Eq.\ref{Eqthetaphi}) \\
\noalign{\smallskip}
{; scattering intensity using HG-phase function $4\pi\,f_I(\theta,g)$ (Eq.~\ref{EqScatPhase}) \hfill}             \\
\noalign{\smallskip}
{\verb|fi = (1.-g^2)/(1.+g^2-2.*g*cos(theta))^1.5|\hfill}  & \hspace{-3cm}; $4\pi\,f_I(\varphi_d,i)$ (Fig.~\ref{Fig1Ring2D}) \\ 
\noalign{\smallskip}
{; polarized intensity using Rayleigh scattering splitting with $p_{\rm max}=1$ (Eqs.~\ref{Eqper},\ref{Eqpar})\hfill}\\
\noalign{\smallskip}
{\verb|fper = fi/(1.+(cos(theta))^2)|\hfill}                  & \hspace{-3cm}; $f_\perp$ or azimuthal intensity        \\
{\verb|fpar = fi*(cos(theta))^2 / (1.+(cos(theta))^2)|\hfill}   & \hspace{-3cm}; $f_\parallel$ or radial intensity  \\
{\verb|fphi = fper-fpar|\hfill}            & \hspace{-3cm}; $4\pi\, f_\phi(\varphi_d,i)$: azimuthal polarization (Fig.~\ref{Fig1Ring2D}) \\
{\verb|fq = fphi*(-cos(2.*phi))|\hfill}   & \hspace{-3cm}; $4\pi\,f_Q(\varphi_d,i)$ for Stokes $Q_{\rm d}$ (Fig.~\ref{Fig2Ring2D}) \\
{\verb|fu = fphi*(-sin(2.*phi))|\hfill}   & \hspace{-3cm}; $4\pi\,f_U(\varphi_d,i)$ for Stokes $U_{\rm d}$ (Fig.~\ref{Fig2Ring2D}) \\
\noalign{\smallskip}
{; disk averaged scattering functions $\langle f(i,g) \rangle$  \hfill} \\
\noalign{\smallskip}
{\verb|fiavg = mean(fi)|\hfill}            & \hspace{-3cm}; intensity $\langle f_I(i,g) \rangle$  (Eq.~\ref{Eqintphir}) \\
{\verb|fphiavg = mean(fphi)|\hfill}      & \hspace{-3cm}; azimuthal polarization $\langle f_\phi(i,g) \rangle$ (Eq.~\ref{Eqintqphi}) \\   
\noalign{\smallskip}
{; normalized quadrant polarization parameters\hfil}  \\
\noalign{\smallskip}
{\verb|qpp = fltarr(5)|\hfil}            & \hspace{-3cm}; initialize quadrant values \\
\noalign{\smallskip}
{; sum-up of relevant $\phi_{xy}(\varphi_d,i)$-points for each quadrant (according to Tab.~\ref{Integration})\hfil}  \\
\noalign{\smallskip}
{\verb|for j=0,3599 do begin|}                             \\
{\verb|if (phi[j] gt -0.25*!pi and phi[j] lt 0.25*!pi)then qpp[0]=qpp[0]+fq[j]/3600.| \hfill} & ; quadrant Q$_{000}$\\
{\verb|if (phi[j] gt 0.       and phi[j] lt 0.50*!pi) then qpp[1]=qpp[1]+fu[j]/3600.| \hfill} & ; quadrant U$_{045}$\\
{\verb|if (phi[j] gt 0.25*!pi and phi[j] lt 0.75*!pi) then qpp[2]=qpp[2]+fq[j]/3600.| \hfill} & ; quadrant Q$_{090}$\\
{\verb|if (phi[j] gt 0.50*!pi and phi[j] lt 1.00*!pi) then qpp[3]=qpp[3]+fu[j]/3600.| \hfill} & ; quadrant U$_{135}$\\
{\verb|if (phi[j] gt 0.75*!pi or phi[j] lt -0.75*!pi) then qpp[4]=qpp[4]+fq[j]/3600.| \hfill} & ; quadrant Q$_{180}$\\
{\verb|endfor|\hfill}  &      \\
{\verb|return|\hfill}  &      \\
{\verb|end|\hfill}     &      \\
\end{tabular} 
\end{table*}
  
\end{appendix}

\end{document}